\documentclass[aps,floats,amssymb,twocolumn,prb]{revtex4}
\usepackage{calc}
\usepackage{psfrag}
\usepackage{graphicx}
\usepackage{color}
\usepackage{bm}

\begin{document}

\title{Paired states at 5/2: PH Pfaffian and particle-hole symmetry breaking}

\author{ L. Antoni\'c$^{1}$, J. Vu\v ci\v cevi\'c$^{2}$, and M. V. Milovanovi\'c$^{2}$}
\affiliation{$^1$ Faculty of Physics, University of Belgrade, 11001 Belgrade, Serbia}
\affiliation{$^2$ Scientific Computing Laboratory, Center for the Study of Complex Systems, Institute of Physics Belgrade, University of Belgrade, Pregrevica 118, 11080 Belgrade, Serbia}

\begin{abstract}
We study Cooper pairing in the Dirac composite fermion (CF) system. The presence of the mass term in the Dirac CF description (which may simulate Landau level mixing) i.e. breaking of particle-hole (PH) symmetry in this system, is a
necessary condition for the existence of a PH Pfaffian-like topological state. In the scope of RPA approximation and
hydrodynamic approach, we find some signatures of  pairing at finite frequencies. Motivated by this insight, we extend our analysis to the case of a different but still Dirac quasi-particle (CF) representation, appropriate in the presence of a mass term, and discuss the likelihood of PH Pfaffian pairing and Pfaffian pairing in general. On the basis of
gauge field effects, we find for small Dirac mass anti-Pfaffian or Pfaffian instability depending on the sign of mass, while for large mass (Landau level mixing), irrespective of its sign, PH Pfaffian-like instability.

\end{abstract}

\maketitle

\section{Introduction}

The fractional quantum Hall effect (FQHE)[\onlinecite{ts}] is a remarkable effect of electrons confined to two dimensions. In the presence of a strong, perpendicular to the plane magnetic field, the phase space of the strongly correlated system
is further confined into Landau levels (LLs) and quantized. At special fillings of LLs, when the system is described
by special ratios of number of electrons per number of flux quanta, highly entangled states of FQHE are
established. In experiments, the effect is seen by measuring fractionally quantized Hall conductance, which stays
constant, at the particular value of fractional filling factor, as magnetic field or density is varied.

The Laughlin state [\onlinecite{la}] with its generalizations describes the effect at odd denominator filling factors. A surprise came
with the experimental detection of FQHE at filling factor 5/2 i.e. half-filling of the second LL (SLL)[\onlinecite{will}]. The Cooper pairing was
invoked to explain the effect.  Assuming spinless (frozen spin) electrons, at half-filling of active (second) LL,
in the regime of experiments, the most natural BCS pairing function in the real space, which can be associated with an
antisymmetric matrix, is a Pfaffian wave function [\onlinecite{mr}]. Thus we expect Cooper pairing due to phase-space
constraints - gauge field effects in a field theory description, in the presence of the repulsive Coulomb interaction.

But the realization of the pairing correlations even for spinless fermions is not unique. We may envision a theoretical construct, an isolated half-filled LL with the exact particle-hole symmetry that can be explored in numerical experiments. The early pairing proposal - Pfaffian or Moore-Read state [\onlinecite{mr}] does not possess the particle-hole symmetry, and, under particle-hole exchange the Pfaffian transforms  into an anti-Pfaffian state [\onlinecite{apf,sapf}]. The Pfaffian and  anti-Pfaffian equally participate in the ground state of the half-filled SLL with Coulomb interaction [\onlinecite{dh}]. The system with the
particle-hole symmetry requires for its field-theoretical description, a special Dirac composite fermion (DCF) representation of constitutive classical electrons  and their strong correlations [\onlinecite{son}]. On the basis of this representation a proposal was
made for a special Pfaffian that respects the particle-hole symmetry, so-called PH Pfaffian [\onlinecite{son}].

Arguments were given in Refs. [\onlinecite{kachru,wc,pair,mish}]  that the PH Pfaffian in the particle-hole symmetric setting i.e.
half-filled LL is an unstable, critical state. Nevertheless, the PH Pfaffian type of pairing seems relevant from the experimental point of view, as argued in Ref. [\onlinecite{zf}], despite LL mixing (absence of the
particle-hole symmetry) and disorder effects.

Inspired by the recent experiment described in Ref. [\onlinecite{exp}] that measured the thermal Hall conductance of the paired state at filling 5/2, and found that the measured value is consistent with the conductance of PH Pfaffian, we would like to check if a PH Pfaffian-like state may be realized in the absence of the PH symmetry i.e. in the presence of LL mixing, but without disorder. The
possibility for PH Pfaffian physics due to disorder effects is considered in Refs. [\onlinecite{zf,wang,mross,biao}], see also [\onlinecite{halp}],
while in [\onlinecite{shs}] a proposal is made that the result of the experiment may be still consistent with the anti-Pfaffian state, due to an insufficient equilibration of edge modes. Very recently, this proposal is criticized in
Ref. [\onlinecite{df}].

In this work we discuss the effective Cooper pairing channel of the system at
half-filling in the scope of the DCF theory with a mass term. The mass term of the DCF theory represents a term that breaks the particle-hole symmetry of electrons confined in a LL and represents a LL mixing.

The paper is organized as follows. In Section II we review arguments for the
criticality of the PH Pfaffian in a particle-hole symmetric setting, and argue
why a symmetry-breaking mass in the DCF theory is necessary to stabilize the
PH Pfaffian. In Section III we discuss the effective Cooper pairing channel in the DCF theory in the presence of a mass term, and recover only some finite frequency pairing correlations. This is followed by a discussion in Section IV,
which uses a different form of the DCF theory to analyze the Cooper pairing
of modified CFs, in which a usual BCS problem emerges from a gauge field description of constraints. The Pfaffian family solutions of the problem are
described and conclusions can be found in Section V.

\section{PH Pfaffian as a critical state in a half-filled Landau level}

\subsection{PH Pfaffian as a critical state in a (particle-hole symmetric) half-filled Landau level}

In this subsection we will review arguments given in Ref. \onlinecite{pair} for the critical nature of the PH Pfaffian
state, and, in addition, relate the PH Pfaffian physics in a half-filled Landau level to the critical behavior and  transition between Pfaffian and anti-Pfaffian [\onlinecite{apf}], and discuss how general the arguments for the critical nature of PH Pfaffian are.

In the following we will denote by a PH Pfaffian state, a FQHE state, at filling factor 1/2 (half-filled Landau level) with Pfaffian ($p$-wave pairing) correlations that is invariant under particle-hole (PH) transformation. The correlations are expected to be in the opposite sense of rotation with respect to the one set by external magnetic field. On the other hand we will denote by $\tilde{PH}$ Pfaffian, a state with Pfaffian correlations in the opposite sense of rotation with respect to the one set by external magnetic field. The $\tilde{PH}$ Pfaffian
 is a generalization of PH Pfaffian and may not have the PH symmetry. The Pfaffian state with the PH symmetry was mathematically defined as a $s$-wave pairing instability of an effective description by Dirac composite fermions at half-filling. That is known in the literature as the Son's proposal for the PH (symmetric) Pfaffian [\onlinecite{son}].

 Based on a mean field analysis, we will argue that PH Pfaffian and its $\tilde{PH}$ Pfaffian extensions, in the presence of the PH symmetry (i.e. in a system with PH symmetric Hamiltonian), describe {\em critical} states and thus
they can not describe gapped topological phase(s) in half-filled LLs.

First we will examine the underlying physics behind the states with so-called negative flux insertion, either macroscopically as in [\onlinecite{t}], or microscopically as in [\onlinecite{zf}], that induces $p$-wave pairing in the opposite sense of the rotation with respect to the one set by the external field. These states can be described as $\tilde{PH}$ Pfaffian states. As long as we are
not sure of the fate of the PH symmetry in these constructions we will consider them as $\tilde{PH}$ Pfaffians.

A negative flux Pfaffian ( a $\tilde{PH}$ Pfaffian) lowest LL (LLL) wave function was introduced in Ref. [\onlinecite{t}] as
\begin{eqnarray}
\Psi_{TJ}& = & P_{LLL} [ {\cal S } \{\prod (z_{i1 }^* - z_{j1}^*)^2 \times \prod (z_{i2}^* - z_{j2}^* )^2 \}  \nonumber \\
&& \times  \prod (z_{k} - z_{l})^3]  , \nonumber \\
 \label{tj}
\end{eqnarray}
where $\{ z_i = x_i + i y_i , i = 1,\ldots, N \}$ are the electron coordinates, we omitted the Gaussian factors, $P_{LLL}$ projects to the lowest LL, and the symmetrizer ${\cal S}$ symmetrizes between two groups, 1 and 2, in which the particles are equally distributed. This is a state with $N_\phi = 3 N_e - 3 - 2 (N_e/2 - 1) = 2 N_e - 1, $ i.e. with a PH symmetric shift. An algebraic procedure introduced in [\onlinecite{mmr}] may be followed to generate possible edge states i.e. sector, if the system is incompressible. Namely the proposition applied in the procedure is that if we consider bulk quasihole coherent state constructions, we can use them to generate edge states of an incompressible state. The method proved successful in the Pfaffian case especially so because, in that case, the edge states can be defined as those that make energy zero subspace of a model interaction for which the ground state - Pfaffian model wave function is also a zero energy state. In the case of the state in (\ref{tj}) we are not aware of the existence of a model interaction, and, furthermore, we do not know if the state is incompressible. Nevertheless, we may examine what well-known, well-motivated quasiparticle bulk constructions can generate as low-momentum states. If the state is incompressible these states we expect will make  the edge sector. The analysis was done in  [\onlinecite{mj}], and the states recovered in this way, under the assumption of the incompressibility, would make counterpropagating Majorana edge branch, together with the charged boson edge branch. The analysis missed a neutral copropagating neutral boson, which states can be described as insertions, in the antiholomorphic  part of the wave function, of holomorphic differences of symmetric polynomials belonging to two groups of particles under symmetrizer. (The antiholomorphic differences i.e. their linearly independent combinations under symmetrization, make the states of the counterpropagating Majorana edge branch.)

Next to the construction in (\ref{tj}) we can consider a $\tilde{PH}$ Pfaffian state:
\begin{equation}
\Psi_{ZF} =  P_{LLL} [ Pf \{ \frac{1}{(z_{i}^* - z_{j}^*)} \}\prod (z_{k} - z_{l})^2],
 \label{zf}
\end{equation}
which was introduced in Ref. [\onlinecite{zf}]. Here
\begin{eqnarray}
Pf\{ \frac{1}{(z_i^* - z_j^*)} \} \sim \nonumber \\
 \sum_P sgn \; P \prod_{i=1}^{N/2} \frac{1}{(z_{P(2 i - 1)}^* - z_{P(2 i)}^*)}.  \label{cPf}
\end{eqnarray}
In this case, an analysis of the edge states can include only antiholomorphic Majorana (neutral fermion) constructions described in the Moore-Read (holomorphic Pfaffian) case in Ref. [\onlinecite{mmr}]. Thus next to the charged boson we have only a single counterpropagating Majorana.

At this stage it is interesting to note that in one of the papers that introduced the anti-Pfaffian physics, in Ref.
[\onlinecite{apf}], two out of three states that may appear at the  transition between Pfaffian and anti-Pfaffian, have the same edge physics as described here for the states in (\ref{tj}) and (\ref{zf}).

In the following we will demonstrate that the states in (\ref{tj}) and (\ref{zf}), that may describe electrons in half-filled LLs, are in fact critical states i.e. gapless and unstable states. For that we will consider a simplified (mean field) version of the Son's theory - the DCF theory [\onlinecite{son}] that describes the Fermi-liquid-like state of Dirac CFs, in a half-filled i.e. PH symmetric Landau level. Thus we consider a massless Dirac fermion, at finite density, with $s$-wave pairing among spinor components, and neglect the presence of the gauge field i.e. its fluctuations around zero value. In the chirality basis i.e. in the basis of Dirac eigenstates without pairing, see Ref. [\onlinecite{pair}] for details, we can express the pairing term which pairs spinor components $a$ and $b$ as
\begin{equation}
\Psi_a ({\bf k})  \Psi_b ( - {\bf k})
=- \frac{1}{2}  \frac{k_+}{k} [\Psi_+ ({\bf k})  \Psi_+ ( - {\bf k}) +   \Psi_- ({\bf k}) \Psi_- (-{\bf k})],
\label{s_into_p}
\end{equation}
where $k \equiv |{\bf k}|$ and $k_+ = k_x + i k_y$. The fermion fields $\Psi_+$ and $ \Psi_-$ represent definite chirality (eigenstates of $\frac{{\vec{\sigma}} \cdot{\vec k}}{k}$) particle (positive energy) and hole (negative energy) states. As the relevant low-energy physics is around finite chemical potential, for the description of the pairing physics we may use the following low-energy, decoupled from higher modes, BCS Hamiltonian,
\begin{eqnarray}
H_{BCS}&=& \sum_{\bf k} (k - \mu) \Psi_+^\dagger ({\bf k}) \Psi_+ ({\bf k}) +  \nonumber \\
&& \sum_{\bf k} \{ \frac{1}{2} \frac{k_+}{k} \Delta_s \Psi_+ ({\bf k})  \Psi_+ ( - {\bf k}) + h.c.\}.
\label{pH}
\end{eqnarray}
We arrived to the usual form of the $p$-wave spinless fermion pairing Hamiltonian as can be found in [\onlinecite{rg}]
except that here we have linearly dispersing fermions, and not fully specified $\Delta_s$ function in the pairing part. With respect to the notation of Ref. [\onlinecite{rg}], the pairing function can be identified as
\begin{equation}
\Delta_{\bf k}^* = - \frac{k_+}{k} \Delta_s .
\label{delta}
\end{equation}
The algebra of the Bogoliubov problem (Ref. [\onlinecite{rg}]) leads to the following expression for the Cooper pair wave function:
\begin{equation}
g_{\bf r}  =  \frac{1}{V} \int d{\bf k} \exp\{i {\bf k}{\bf r}\} \frac{-(E_{\bf k} - \xi_{\bf k})}{\Delta_{\bf k}^*},
\end{equation}
where $ \xi_{\bf k} = k - \mu$, and $E_{\bf k}^2 = \xi_{\bf k}^2 + |\Delta_{\bf k}|^2$. We are interested in the long-distance behavior, and thus the behavior of $\Delta_{\bf k}$ for small momenta around $ k = 0$. For $\mu > 0$, finite chemical potential and density of the system, that we consider here, the long distance behavior is determined by the behavior of $\Delta_{\bf k}$ i.e. $\Delta_s$ (see Eq. (\ref{delta})) for small ${\bf k}$.  In the small $k$ limit
we are motivated to consider two cases:
\begin{equation}
(a) \lim_{k \rightarrow 0} \Delta_s \rightarrow {\rm Const.}, \;\;{\rm when}\;\; g_{\bf r} \sim \frac{1}{z |z|}
\label{a}
\end{equation}
and
\begin{equation}
(b) \lim_{k \rightarrow 0} \Delta_s \sim k \equiv |{\bf k}|,\;\; {\rm when}\;\; g_{\bf r} \sim \frac{1}{z}.
\label{b}
\end{equation}
For the usual choice of the direction of the magnetic field $\vec{B} =  B \; \hat{e}_z, B < 0$  (instead of $B > 0$
as is implicit in the DCF theory because the density of Dirac CFs is prpotional to $B$) we would have  in (\ref{a}) and (\ref{b}), instead of $z$, in fact $z^*$. Thus, in the long-distance limit, when we can neglect the projection to a definite LL, the case (\ref{b}) corresponds to wave functions in (\ref{tj}) and (\ref{zf}), because
\begin{eqnarray}
&&{\cal S } \{\prod (z_{i1 }^* - z_{j1}^*)^2 \times \prod (z_{i2}^* - z_{j2}^* )^2 \} =  \nonumber \\
&&  Pf\{ \frac{1}{(z_i^* - z_j^*)} \} \times \prod (z_{k}^* - z_{l}^*),
 \label{eq}
\end{eqnarray}
and thus the pairing with the Cooper pair wave function, $g_{\bf r} \sim \frac{1}{z^*}$, is present in both wave functions.
Thus, we can conclude, in order to reproduce the pairing encoded in wave functions, (\ref{tj}) and (\ref{zf}), we need non-analytic behavior in the small k limit, $\lim_{k \rightarrow 0} \Delta_s \sim k \equiv |{\bf k}|$.

Thus, for the states in Eq.(\ref{tj}) and Eq.(\ref{zf}), we cannot have a Landau-Ginzburg type of description (together with the fermionic part, see below), as we would expect to have for a well-defined, stable pairing phase. A question may be raised, whether in this argument for the critical nature of these states, we are allowed to use the region around $|{\bf k}| = 0$ to describe the pairing instabilities of Dirac CFs, with the understanding that the Dirac description is only well-defined near $|{\bf k}| = k_F$. But a complete theory (description) of a pairing instability involves both regions; the one around $|{\bf k}| = k_F$ and the one around $|{\bf k}| = 0$, associated with the long-wavelength, low-energy description with a bosonic variable (an order parameter - $\Delta_{\bf k}$ in the mean-field description) associated with the pairing.

On the other hand, the introduction of an analytical $s$-wave pairing in the DCF theory would lead to the following Lagrangian density,
\begin{eqnarray}
{\cal L}& = & i \bar{\chi} \gamma^\mu (\partial_\mu + i a_\mu) \chi + (i g \tilde{\Delta}_s ({\bf r}) \chi \sigma_y \chi + h.c.) \nonumber \\
&& + |(\partial_\mu - 2 i a_\mu) \tilde{\Delta}_s|^2 - u|\tilde{\Delta}_s|^2 - \frac{v}{2} |\tilde{\Delta}_s|^4 .
\end{eqnarray}
The theory is invariant under $CP$ (charge conjugation + parity) transformation
\begin{equation}
CP \chi ({\bf r}) (CP)^{-1} = \sigma_x \chi (\bf r'),
\end{equation}
where ${\bf r} = (x, y)$ and ${\bf r'} = (x, - y)$, though the pairing term up to a gauge transformation. This invariance corresponds to the invariance under the PH transformation of real electrons [\onlinecite{son}]; the introduced $s$-wave
pairing constitutes the Son's proposal for the PH Pfaffian. With the usual $s$-wave pairing behavior, $\lim_{k \rightarrow 0} \Delta_s \rightarrow {\rm Const.}$, and neglecting the influence of gauge field, as before, we can arrive to the following, characteristic long-distance behavior,
\begin{equation}
 g_{\bf r} \sim \frac{1}{z |z|} ({\rm i.e.}  \frac{1}{z^* |z|}),
\end{equation}
which should enter the Pfaffian part of the PH Pfaffian,
\begin{equation}
\Psi_{PH} =  P_{LLL} [ Pf \{ \frac{1}{(z_{i}^* - z_{j}^*)|z_{i} - z_{j}|} \}\prod (z_{k} - z_{l})^2].
 \label{ph}
\end{equation}
We will list two reasons why this state can be considered only as a gapless (critical) state:
(a) If we attempt to generate edge Majorana sector using $\Psi_{PH}$, and the method of Ref. [\onlinecite{mmr}], we will not be able to separate (bulk) charge modes from the usual Majorana counterpropagating edge modes, because this is possible only for the peculiar form of the Cooper pair wave function with $ g_{\bf r} \sim \frac{1}{z^*}$.  (b) The universal, long-distance behavior $ g_{\bf r} \sim \frac{1}{z^*|z|}$ is also a characteristic behavior of the critical system of classical (non-relativistic) fermions at the transition from weak to strong coupling as described in [\onlinecite{rg}].

Though we commented in the beginning that we are applying mean field approach and neglecting gauge field fluctuations, our approach is in fact quite general, and can reach conclusions that are not biased. Namely, once we assume PH Pfaffian pairing instabilities, by the very assumption of the pairing order parameters of underlying Dirac composite fermions, we expect, due to the Anderson-Higgs mechanism, that gauge field (that couples to the fermions) will be expelled from the low-energy physics, and therefore our assumption and approach concerning the nature of PH Pfaffian is justified.

Thus the state with a manifest PH symmetry, the PH Pfaffian in the half-filled LL can be only a critical state. This state may correspond to the (third) state that characterizes  the transition between Pfaffian and anti-Pfaffian state in Ref. [\onlinecite{apf}]. According to the analysis of Ref. [\onlinecite{apf}]
we may expect that this state is the lowest energy state at the transition between Pfaffian and anti-Pfaffian state
i.e. a ``real" critical state, while other two states, in (\ref{tj}) and (\ref{zf}), may represent excited states at the point of the transition with an exact particle-hole symmetry.
(The two states in (\ref{tj}) and (\ref{zf}), correspond to the other two states of the same reference, due to their edge spectrum.)

\subsection{Particle-hole symmetry breaking  and the criticality of PH Pfaffian}

It is interesting to introduce  a mass i.e. a particle-hole symmetry breaking term in the previously discussed description of the pairing instabilities in a fixed Landau level. We will assume the analytic ($\lim_{k \rightarrow 0} \Delta_s \sim {\rm Const})$ description, discussed in the previous subsection. At the particle-hole symmetric point $ m ({\rm mass}) = 0$. Away from this point, for $\Delta_s = 0$, we have a simple Dirac description of  the Hamiltonian with the following
$2 \times 2$ matrix,
\begin{equation}
H = \left[
  \begin{array}{cc}
   m & k_- \\
   k_+ & - m \\
  \end{array}
\right],
\end{equation}
for the following choice for gamma matrices $ \gamma^0 = \sigma_3, \gamma^1 = i \sigma_2,$ and $\gamma^2 = - i \sigma_1$, and we set
 the Dirac velocity , $v_F = 1$. In this case,
\begin{equation}
\Psi_+ ({\bf k}) = \frac{1}{\sqrt{2 E (E + m)}} [(m + E) \Psi_a ({\bf k}) + k_- \Psi_b ({\bf k})],
\end{equation}
and
\begin{equation}
\Psi_- ({\bf k}) = \frac{1}{\sqrt{2 E (E - m)}} [(m - E) \Psi_a ({\bf k}) + k_- \Psi_b ({\bf k})],
\end{equation}
where $ E \equiv \sqrt{|{\bf k}|^2 + m^2}$. We find that
\begin{eqnarray}
 \Psi_a ({\bf k})  \Psi_b ( - {\bf k})
& = &  - \frac{k_+}{2 E}   \Psi_+ ({\bf k})  \Psi_+ ( - {\bf k}) \nonumber \\
&& -  \frac{m}{ E} \frac{k_+}{|{\bf k}|}  \Psi_+ ({\bf k}) \Psi_- (-{\bf k}) \nonumber \\
&&  + \frac{k_+}{2 E}   \Psi_- ({\bf k})  \Psi_- ( - {\bf k}). \label{pairing_s_with_m}
\end{eqnarray}
We immediately see that in this case, with respect to the Eq.(\ref{s_into_p}), we do not have non-analiticity for
$ {\bf k} = 0$ (if $\Delta_s$ is a constant in that limit). Thus, for $ m \neq 0$, we have a description similar to the ordinary $p$-wave in Ref. [\onlinecite{rg}], that reproduces Pfaffian pairing in the opposite direction with respect to the one set by the external magnetic field, as discussed in the previous subsection, but with a mixing term $ \sim
\Psi_+ ({\bf k}) \Psi_- (-{\bf k})$.

A straightforward solution of the Bogoliubov - BCS problem gives  a BCS ground state where $\Psi_+ $ degrees of freedom pair as $ g_{\bf r} \sim \frac{1}{z}$, while interband correlations are described with $ g_{\bf r} \sim \frac{1}{z |z|}$. Thus the implied ground state wave function in the effective long-wavelength description is
\begin{equation}
\Psi_{PHL} =   Pf \{ \frac{1}{(z_{i}^* - z_{j}^*)} \}\prod (z_{k} - z_{l})^2 .
 \label{phl}
\end{equation}
This leads to the conclusion that the particle-hole symmetry breaking mass term may stabilize the PH Pfaffian -like state in (\ref{phl}). This is a very interesting, counterintuitive conclusion, which was originally suggested in Ref.
[\onlinecite{pair}], in the context of singlet and triplet pairings of spinor components, in the presence of a mass term. Here we showed that the same conclusion can be reached by considering only the $s$-wave (singlet) pairing (Eq. (\ref{pairing_s_with_m})) in  the presence of a mass term.

Although this simple scenario seems quite plausible, the numerical investigations of the second Landau level (SLL) (for which we expect that is dominated by the Pfaffian physics) imply that the physics around the particle-hole symmetric point is dominated by a non-universal influence of the short-range part of the Coulomb interaction, which is hard to capture by field theoretical means. Namely the investigation in Ref. [\onlinecite{dh}] clearly shows the (Schroedinger cat) mixing of Pfaffian and anti-Pfaffian at the particle-hole symmetric point, and their relevance for the nearby physics. The most recent investigation in Ref. [\onlinecite{mish}] points out that the state in Eq. (\ref{zf}) is likely an excited state in the half-filled SLL (compare with our identification above), and has very high overlap with composite Fermi liquid (CFL) wave function [\onlinecite{rere,rh}]. Thus, although the DCF theory seems a very good description of the half-filled LLL, it has to be modified to capture the non-universal physics in the SLL. But, by modifying the Coulomb interaction in the SLL, one can stabilize the state in (\ref{zf}) or the CFL state  [\onlinecite{mish}]. Thus a relevant question may be whether a mass term in the DCF theory may induce pairing irrespective of the details of the projected to a LL Coulomb interaction.

Therefore we are motivated to study the DCF theory in the presence of a mass term in order to see if this may induce pairing correlations and a pairing instability of the PH Pfaffian kind (Eq.(\ref{phl})).

\section{The Dirac composite fermion theory with a particle-hole symmetry breaking term}

In this section we will consider the usual formulation of the DCF theory, with a mass term, in the RPA approximation in order to find the effective Cooper channel and examine pairing correlations. A different formulation of the same theory we will discuss in the next section.

\subsection{The Dirac composite fermion theory with a particle-hole symmetry breaking term - an introduction}
We start by examining the DCF theory in the presence of PH symmetry breaking mass term.The generalized Lagrangian
can be found in Ref. [\onlinecite{pot}], and is given by the following expression,
\begin{eqnarray}
{\cal L}& = & i \bar{\psi} \gamma^\mu (\partial_\mu + i a_\mu) \psi + \frac{1}{4 \pi} a dA  + \frac{1}{8 \pi} A dA  \nonumber \\
&&  - m  \bar{\psi} \psi , \label{mL}
\end{eqnarray}
where $\chi $ is the Dirac CF field, $a_\mu $ is an emergent $U(1)$ gauge field, the Chern-Simons terms are abbreviated
as $\epsilon^{\mu \nu \lambda} A_\mu \partial_\nu A_\lambda \equiv A \partial A,\; \mu = 0, x, y$, and we have omitted the Coulomb interaction and higher order terms.
As a consequence we have the following equations, by differentiating with resect to $A_\mu$,
\begin{equation}
J_e = \frac{dA}{4 \pi} + \frac{da}{4 \pi},
\label{ej}
\end{equation}
where $J_e$ is the electron density-current, and, by differentiating with resect to $a_\mu$,
\begin{equation}
J_\psi = \frac{dA}{4 \pi},
\label{fj}
\end{equation}
where $J_\psi$ is the Dirac composite fermion density-current. As discussed in [\onlinecite{pot}] the transport coefficients like Hall conductance can be found from the implied form of currents,
\begin{equation}
{\bf j}_e = \frac{1}{4 \pi} \hat{\epsilon} ({\bf E} - {\bf e}),
\end{equation}
and
\begin{equation}
{\bf j}_\psi = \frac{1}{4 \pi} \hat{\epsilon} {\bf E},
\end{equation}
where $\hat{\epsilon}$ is the unit antisymmetric tensor with components $\epsilon^{ii} = 0, \epsilon^{xy} = - \epsilon^{yx} = 1$,
together with the relationship that we have to extract from the theory,
\begin{equation}
{\bf j}_\psi = \frac{1}{4 \pi} \hat{\sigma}_D {\bf e},
\end{equation}
where $\hat{\sigma}_D$ represents the Dirac composite fermion conductance tensor. To find $\hat{\sigma}_D$
we need to find the polarization tensor $\Pi_{\mu \nu}$,
\begin{equation}
j^\psi_\mu = \Pi_{\mu \nu} \; a^\nu ,
\end{equation}
which may be identified in the RPA treatment of the theory: the expansion of the effective action to second order in
$a^\nu $, after the integration of fermion fields in the functional formalism, or directly calculating
\begin{equation}
\Pi^{\mu \nu} = - i tr[ \gamma^\mu S_F (x, y)\gamma^\nu S_F (y, x)], \label{defPi}
\end{equation}
where
\begin{equation}
i S_F(x, y) = \langle T[\psi(x) \bar{\psi}(y)]\rangle , \label{fer_prop_def}
\end{equation}
i.e. the composite fermion propagator ($T$ is the time ordering and the expectation value is with respect to the ground state of non-interacting fermions). In calculating (\ref{defPi}) we encounter (ultra-violet) divergences, which come from the presence of the infinite sea of negative energy solutions. There are two ways to regularize the theory: (a) dimensional and (b) Pauli-Villars regularization. The physical meaning of these two possibilities, when considering Berry curvature contributions of the positive and negative energy band to the Hall conductance, is  that in the dimensional regularization we combine (add) the contributions, while in the Pauli-Villars regularization we consider the contribution only  from the positive band. (For the Berry curvature contributions of the two bands see Ref. [\onlinecite{pot}].) The dimensional regularization at the neutrality point ($\mu$(chemical potential) = 0), and in the presence of a mass gives an unphysical prediction for the Hall conductance ($= \frac{1}{2} \frac{e^2}{h}$)  i.e. a half integral quantum Hall effect of noninteracting fermions. But at a finite chemical potential, and in the absence of mass the Hall conductance is zero. This result or consequence is at the basis of the DCF theory (that is defined at a finite chemical potential), which results in the precise value of the Hall conductance of electrons $= \frac{1}{2} \frac{e^2}{h}$, dictated by the  particle-hole symmetry, even in the presence of disorder. Thus the dimensional regularization is the assumed regularization in the DCF theory. On the other hand, the Pauli-Villars regularization gives a so-called parity anomaly, a half of the unit of the Hall conductance even in the absence of a mass. Later we will explore the role of the Pauli-Villars regularization, when we consider a smooth connection between the DCF theory and the HLR theory [\onlinecite{rev}]. Thus this type of regularization is important when we switch the quasiparticle representation from the one based on the Read's construction [\onlinecite{read}] (the DCF theory) to the one based on the usual Chern-Simons construction [\onlinecite{zhang}] (the HLR theory [\onlinecite{hlr}]).

We obtained the polarization tensor $\Pi^{\mu \nu}({\bf k},\omega)$, in the hydrodynamic approximation i.e. when $ |{\bf k}| \ll k_F$, in the presence of the mass term. The results can be found in Appendix A.

\subsection{The Cooper channel in the Dirac composite fermion theory with a particle-hole symmetry breaking term}

In this subsection we will extend the approach applied in Refs. [\onlinecite{kachru,wc}] to the case with the particle-hole symmetry breaking mass term. Namely, in Ref. [\onlinecite{kachru}], in order to study the possibilities for pairing within  the DCF theory, the Coulomb interaction was considered as an additional term in the theory
described by Eq.(\ref{mL}) with $m = 0$. An effective Coulomb interaction was found by a projection of  fermion operators to the low-energy sector around $|{\bf k}| = k_F$ of positive energy solutions.
The BCS interaction or Cooper channel of the effective interaction for the $p$-wave (PH Pfaffian) pairing was found to be repulsive and in no way conducive for the pairing. The investigations in [\onlinecite{kachru,wc}] included, at the RPA level, modifications of the effective interaction due to the fluctuations of the gauge field ($a_\mu$), but the conclusion was the same. In this subsection we  would like to find out the effective Cooper channel, within the RPA approach, in the presence of a mass term.

To investigate the possibility for a stable pairing phase, we will look for the expression of the effective interaction in the imaginary time formalism, but fix $T({\rm temperature}) = 0$. In the Euclidean space-time we have
\begin{eqnarray}
{\cal L}_E & = &  \bar{\psi}_E \gamma^0 (\partial_\tau +  a_E^0) \psi_E  + \bar{\psi}_E (-i {\bf \gamma}) ({\bf \nabla} + i{\bf a}_E) \psi_E  \nonumber \\
&& - \mu \bar{\psi}_E \gamma^0 \psi_E   + m  \bar{\psi}_E \psi_E , \label{EuclmL}
\end{eqnarray}
where we consider the situation with a constant magnetic field and thus, the fermi system at a finite chemical potential $\mu$. We can get (\ref{EuclmL}) from (\ref{mL}) by a naive analytical continuation $ \tau = i t$. We may introduce Euclidean gamma matrices $\gamma_E = \gamma_0$ and $\vec{\gamma}_E = (-i) \vec{\gamma}$, but, to make an easier contact with previous calculations and literature, we will keep a Minkowski set: $\gamma_0 = \sigma_3 , \vec{\gamma} = i \vec{\sigma}$. We introduce the gauge field propagator by the functional integration over fermionic degrees of freedom,
\begin{eqnarray}
\int {\cal D} \bar{\psi}_E  {\cal D} \psi_E  \exp\{- \int d\tau d{\bf x} [{\cal L}_E]\} = \nonumber \\
\exp\{- \int dx \int dy \frac{\bar{{\cal D}}_{\mu \nu}^{-1}}{2}(x - y) a^\mu (x) a^\nu (y) \}.
\end{eqnarray}
Therefore
\begin{equation}
\bar{{\cal D}}_{\mu \nu}^{-1}(x - y) = tr[ \gamma_\mu G_E (x, y)\gamma_\nu G_E (y, x)],
\label{invDbar}
\end{equation}
where
\begin{equation}
 G_E(x, y) =  - \langle T_\tau[\psi_E(x) \bar{\psi}_E(y)]\rangle ,
\end{equation}
and $x$ and $y$ are points in the Euclidean space-time. We present explicit expressions for $\bar{{\cal D}}_{\mu \nu}^{-1}$, in the hydrodynamic approximation, in Appendix B.

With the addition of the Coulomb interaction to ${\cal L}_E $ (Eq.(\ref{EuclmL})), its contribution to the propagator of the vector potential $a_i , i = x, y$ can be found by considering
\begin{equation}
\delta {\cal L}_E = \frac{\int d{\bf q} \int d\omega}{(2 \pi)^3} \frac{1}{2} (\frac{{\bf q} \times {\bf{a}}(- {\bf q})}{4 \pi})  \frac{2 \pi e^2}{\epsilon_r q} (\frac{{\bf q} \times {\bf{a}}({\bf q})}{4 \pi}).
\end{equation}
To get the effective interaction among fermions, at the RPA level, we integrate out gauge fields in the transverse gauge $(\vec{\nabla}\cdot\vec{a} = 0)$. If we define the fermion density-current as
\begin{equation}
\frac{\delta {\cal L}_E}{\delta a^\mu_E} = {\cal J}^\mu ,
\end{equation}
for the effective 4-fermion interaction we get:
\begin{equation}
V_{int} (x - y) = - \frac{1}{2} {\cal D}_{\mu \nu}(x - y) {\cal J}^\mu (x) {\cal J}^\mu (y),
\end{equation}
where by ${\cal D}_{\mu \nu}$ we denoted the gauge-field propagator with the Coulomb interaction contribution. The propagator ${\cal D}_{\mu \nu}$ can be found in Appendix B.

To find the second quantized expressions for the currents and the interaction we use the following expansion for the fermionic operator,
\begin{equation}
\Psi_E ({\bf x}) = \sum_{\bf k}  \left[
  \begin{array}{c}
i  k_-  \\
 E - m  \\
  \end{array}
\right]  \frac{1}{\sqrt{2 E (E - m)}} \exp\{i {\bf k}{\bf x}\} \;\;  c_{{\bf k}} + \ldots ,
\end{equation}
where we did not write the negative energy contribution.  To describe the Cooper channel we project all momenta to the Fermi circle i.e. $k_\pm = k_x \pm i k_y = k_F \exp\{\pm i \theta\}$.
Starting from the defining expression
\begin{equation}
{\cal J}_0 (x) = \bar{\Psi}_E (x) \gamma_0 \Psi_E (x) = \Psi_E^\dagger (x) \Psi_E (x),
\end{equation}
we find an effective expression for the density operator,
\begin{eqnarray}
&& {\cal J}_0 ({\bf k}_1 - {\bf k}_2 ) =   \nonumber \\
&& \exp\{i \frac{(\theta_2 - \theta_1)}{2}\} \times \nonumber \\
&& [ \cos\{ \frac{(\theta_2 - \theta_1)}{2}\} + i \frac{m}{\mu} \sin\{ \frac{(\theta_2 - \theta_1)}{2}\} ] c_{{\bf k}_2}^\dagger  c_{{\bf k}_1}.
\end{eqnarray}

Defining the transverse part of the current operator by
\begin{equation}
{\cal J}_T ({\bf k}_1 - {\bf k}_2 ) = i \hat{q}\times \bar{\Psi}_E ({\bf k}_1) \vec{\gamma} \Psi_E ({\bf k}_2),
\end{equation}
where $ {\bf q} = {\bf k}_1 - {\bf k}_2 $  $ \hat{q} = \frac{{\bf q}}{|{\bf q}|},$
we find  the effective expression to be
\begin{eqnarray}
&& {\cal J}_T ({\bf k}_1 - {\bf k}_2 ) =   \nonumber \\
&& i \frac{k_F}{\mu} \exp\{i \frac{(\theta_2 - \theta_1)}{2}\} \times
\frac{ \sin\{ \frac{(\theta_2 - \theta_1)}{2}\}}{|\sin\{ \frac{(\theta_2 - \theta_1)}{2}\}|}  c_{{\bf k}_2}^\dagger  c_{{\bf k}_1}.
\end{eqnarray}
Note the presence of the sine function which ensures the hermiticity of the operator $({\cal J}_T^\dagger ({\bf q}) = {\cal J}_T (-{\bf q})$). This part is missing in Ref. [\onlinecite{kachru}].

If we denote the components of  ${\cal D}^{-1}$ by
\begin{equation}
{\cal D}^{-1}({\bf q}, \omega) = \left[
  \begin{array}{cc}
   \hat{\Pi}_{00}  &  \hat{\Pi}_{0T} \\
   \hat{\Pi}_{0T} &  \hat{\Pi}_{TT}\\
  \end{array}
\right],
\end{equation}
the effective interaction potential is
\begin{eqnarray}
&& V_{int}({\bf q}, \omega ) =   \nonumber \\
&& - \frac{1}{[\hat{\Pi}_{00} \hat{\Pi}_{TT}  -(\hat{\Pi}_{0T})^2 ]}  \times \nonumber \\
&& [  \hat{\Pi}_{TT} {\cal J}_0 (-{\bf q}) {\cal J}_0 ({\bf q}) + \hat{\Pi}_{00} {\cal J}_T (-{\bf q}) {\cal J}_T ({\bf q})-  \nonumber \\
&& 2  \hat{\Pi}_{0T} {\cal J}_0 (-{\bf q}) {\cal J}_T ({\bf q})]. \label{comint}
\end{eqnarray}
We can find the effective Cooper channel, by taking expressions for the components of the gauge field propagator (\ref{b3}-\ref{b5}) with $k_0 = i \omega $ where $\omega$ is real, inserting the components in the expression for the effective interaction in (\ref{comint}), and choosing momenta to describe a Cooper pair scattering.

Before a closer look at the effective Cooper channel, we may note that always $ \hat{\Pi}_{00}(i \omega, {\bf k}) < 0 $ and $ \hat{\Pi}_{TT}(i \omega, {\bf k}) > 0 $.

In the static limit $(\omega = 0)$ we have $ \hat{\Pi}_{00}(0, {\bf k})  = - \frac{\mu}{2 \pi}$ , $ \hat{\Pi}_{TT}(0, {\bf k}) = \alpha \frac{|{\bf k}|}{(4 \pi)^2} $, and $ \hat{\Pi}_{0T}(0, {\bf k}) = 0 $. In this case the Cooper channel is
\begin{eqnarray}
&& V^{\rm Cooper}_{\rm int}({\bf q} ={\bf p} - {\bf k}, \omega = 0) = \nonumber \\
&& \{- \frac{1}{\hat{\Pi}_{00}(0, {\bf q})}
 [\cos \frac{(\theta_{\bf k} - \theta_{\bf p})}{2} + i \frac{m}{\mu} \sin \frac{(\theta_{\bf k} - \theta_{\bf p})}{2}]^2 \nonumber \\
&& + \frac{1}{\hat{\Pi}_{TT}(0, {\bf q})} (\frac{k_F}{\mu})^2 \}
 \exp\{i (\theta_{\bf k} - \theta_{\bf p}) \} c_{\bf k}^\dagger c_{\bf p}c_{-{\bf k}}^\dagger c_{-{\bf p}} \nonumber \\
&&
\end{eqnarray}
In the scope of the hydrodynamic approximation i.e. $\theta_{\bf k} \approx \theta_{\bf p} $ we find repulsive behavior and no cause for a Cooper  instability even in the massive case.

The second limit we want to consider is a finite frequency  limit $ \omega \gg \alpha_F |{\bf k}|$. We have $ \hat{\Pi}_{00}(\omega, {\bf k})  \approx - \frac{\mu}{4 \pi} \frac{(\alpha_F |{\bf k}|)^2}{\omega^2} \sim 0 $, $ \hat{\Pi}_{TT}(\omega, {\bf k})  \approx  \frac{\mu}{4 \pi} + \alpha \frac{|{\bf k}|}{(4 \pi)^2} $, and
$ \hat{\Pi}_{0T}(\omega, {\bf k})  \approx - \frac{1}{4 \pi}  \frac{m}{\mu} |{\bf k}|$.
The effective Cooper channel can be described with two terms, (a) density-current part,
\begin{eqnarray}
&& V^{\rm Cooper}_{\rho J}({\bf q} = {\bf p} - {\bf k}, \omega) = \nonumber \\
&& \frac{4 \pi}{m} i \frac{\sin \frac{(\theta_{\bf k} - \theta_{\bf p})}{2}}{|\sin \frac{(\theta_{\bf k} - \theta_{\bf p})}{2}|^2}
 [\cos \frac{(\theta_{\bf k} - \theta_{\bf p})}{2} + i \frac{m}{\mu} \sin \frac{(\theta_{\bf k} - \theta_{\bf p})}{2}] \nonumber \\
&& \times
 \exp\{i (\theta_{\bf k} - \theta_{\bf p}) \} c_{\bf k}^\dagger c_{\bf p}c_{-{\bf k}}^\dagger c_{-{\bf p}} \nonumber \\
&&
\end{eqnarray}
(a) density-density part,
\begin{eqnarray}
&& V^{\rm Cooper}_{\rho \rho}({\bf q} = {\bf p} - {\bf k}, \omega) = \nonumber \\
&& (\frac{4 \pi \mu}{m})^2
\frac{\frac{k_F^2}{4 \pi \mu} + \frac{\alpha}{(4 \pi)^2} (2 k_F) |\sin \frac{(\theta_{\bf k} - \theta_{\bf p})}{2}|}{(2 k_F)^2|\sin \frac{(\theta_{\bf k} - \theta_{\bf p})}{2}|^2}
\nonumber \\
&&
[\cos \frac{(\theta_{\bf k} - \theta_{\bf p})}{2} + i \frac{m}{\mu} \sin \frac{(\theta_{\bf k} - \theta_{\bf p})}{2}]^2 \nonumber \\
&& \times
 \exp\{i (\theta_{\bf k} - \theta_{\bf p}) \} c_{\bf k}^\dagger c_{\bf p}c_{-{\bf k}}^\dagger c_{-{\bf p}} \nonumber \\
&&
\end{eqnarray}
In the density-density part we have extremely singular repulsive interaction present at finite frequencies i.e. a repulsive singular interaction that describes the physics of
excited states. We do not see any cause for a real Cooper instability, except that in the density-current part we can recognize some pairing correlations. This motivates a search for a different quasiparticle representation in which the pairing correlations may be better captured  and exposed.

\section{Pairing correlations within a different quasi-particle representation}

In this section we will consider a different formulation of the DCF theory with a mass term. This will enable us that on the level of equations of motion deduce the effective Cooper channel of different Dirac quasiparticle  from the one
discussed in the preceding section. The channel, derived from purely gauge field effects, supports Pfaffian family
instabilities, and we will examine ensuing phase diagram as a function of  the Dirac mass.
\\

We may also consider the addition of the mass term to the Dirac composite fermion theory by adopting the following form of  the Lagrangian  [\onlinecite{rev}],
\begin{eqnarray}
{\cal L}& = & i \bar{\chi} \gamma^\mu (\partial_\mu + i a_\mu) \chi + \frac{1}{4 \pi} a dA  + \frac{1}{8 \pi} A dA  \nonumber \\
&&  - \frac{m}{|m|} \frac{1}{8 \pi} a da - m  \bar{\chi} \chi . \label{impL}
\end{eqnarray}
Note the presence of the Chern-Simons term for gauge field $a^\mu$. In this case (to recover the identical results for the response with respect to the previous formulation) we have to adopt the Pauli-Villars way of regularizing the theory. Why we discuss this, to say, a redundant formulation? It is important to notice that with a simple redefinition of the gauge field in (\ref{impL}), and in the large mass limit, we can recover the HLR or anti-HLR [\onlinecite{bmf}] theory depending on the sign of mass [\onlinecite{rev}]. See Appendix C for details.

By differentiating with resect to $A_\mu$ the Lagrangian density in (\ref{impL}) we get,
\begin{equation}
J_e = \frac{dA}{4 \pi} + \frac{da}{4 \pi},
\label{edj}
\end{equation}
where $J_e$ is the electron density-current, and, by differentiating with resect to $a_\mu$,
\begin{equation}
J_\chi = \frac{dA}{4 \pi} -  \frac{m}{|m|} \frac{da}{4 \pi},
\label{fdj}
\end{equation}
where $J_\chi$ is the Dirac composite fermion density-current. Thus the Dirac composite fermion density-current in this case is determined, on the classical level, by the fluctuations of the gauge field, just as in usual Chern-Simons theories [\onlinecite{zhang,hlr}]. These theories are based on the quasi-particle (composite fermion) constructions via ``flux tube" - unitary transformations of the original electrons, and not with ``vortex" - Laughlin quasihole constructions [\onlinecite{read}] of quasiparticles. The usual Chern-Simons theories (at the RPA level) can recover the Jastrow-Laughlin correlations [\onlinecite{zhang}] as a part of magneto-plasmon (cyclotron energy) dynamics, while in the ``vortex" constructions they are, in a way, frozen and built in quasi-particles. This distinction may be important when discussing the presence of pairing correlations. In the DCF limit, at the RPA level, both formulations give the same response, because they describe the response of ``vortex" construction, using different regularization schemes. But, at the classical level (Eq.(\ref{fj}) and Eq.(\ref{fdj})), their predictions may differ, because the distinction between the quasiparticle perspectives is preserved. Thus, although, at the RPA level, we find the absence of pairing correlations for ``vortex" constructions (Section III), the fact that some pairing correlations are present in the high-energy i.e. high-frequency sector in the density-current part of the interaction, gives us an expectation that by adopting different quasi-particle representation, we may recover the pairing correlations in the low-frequency or static limit.

The problem, as described by Eq. (\ref{fdj}), is formally identical to the  problem discussed in Ref. [\onlinecite{cai}] in the context of the graphene Dirac electrons in FQHE regime.  Following the analysis of Ref. [\onlinecite{cai}], for a fixed valley, definite spin, Dirac electrons, of the gauge field $a^\mu$ induced interaction between current and density of Dirac particles in the presence of a mass term, we can arrive at the effective form of the pairing channel, Eq. (25) in Ref. [\onlinecite{cai}].

Let's discuss the details that lead to the $p$-wave pairing channel or attractive interaction for a definite, negative sign of the mass, $m < 0$,

The equality in Eq.(\ref{fdj}) will lead to the following integral expression for the gauge field $a^\mu$, $a = a_x + i a_y$,
\begin{equation}
a({\bf r}) = 2 \int d{\bf r}' i \frac{z - z'}{|{\bf r} - {\bf r}'|^2} \delta \rho_\chi ({\bf r}'), \label{a}
\end{equation}
where $ \delta \rho_\chi ({\bf r}')$ represents fermion density with respect to the constant value given by fixed strength of the external magnetic field that we assume.
In the following we will analyze the statistical interaction defined as the one between the current of Dirac fermions and the field $a_i ; i =x, y$
\begin{equation}
V_{st} = \bar{\chi} \gamma^i a_i \chi .
\end{equation}
We work in the following representation of $\gamma$ matrices:
\begin{equation}
\gamma^0 = \sigma_3 , \,\,\, \gamma^1 = i \sigma_2 , \,\,\, \gamma^2 = - i \sigma_1.
\end{equation}
In this representation we have the following expression for the statistical interaction:
\begin{eqnarray}
V_{st} &=& - i 2 \int d{\bf r}' \delta \rho_\chi ({\bf r}') \times  \nonumber \\
&& \chi^\dagger ({\bf r}) \left[
  \begin{array}{cc}
   0 &  \frac{\bar{z} - \bar{z}'}{|{\bf r} - {\bf r}'|^2} \\
   - \frac{z - z'}{|{\bf r} - {\bf r}'|^2} & 0 \\
  \end{array}
\right] \chi({\bf r}),
\end{eqnarray}
and $ \delta \rho_\chi ({\bf r}') = \chi^\dagger ({\bf r}')  \chi ({\bf r}') - \bar{\rho}$, where $\bar{\rho}$ is a constant (external flux density). The constant part gives no contribution to $V_{st}$.

On the other hand the presence of the mass term  in the Dirac system leads to the following eigenproblem,
\begin{equation}
 \left[
  \begin{array}{cc}
   m - \epsilon & k_- \\
   k_+ & - m - \epsilon \\
  \end{array}
\right] \chi({\bf k}) = 0,
\end{equation}
where positive eigenvalue $ \epsilon = \sqrt{|{\bf k}|^2 + m^2} \equiv E_{\bf k}$, corresponds  to the following eigenstate,
\begin{equation}
\chi_E = \left[
  \begin{array}{c}
   m + E_{\bf k} \\
   k_+ \\
  \end{array}
\right]  \frac{1}{\sqrt{2 E_{\bf k} (E_{\bf k} + m)}}.
\end{equation}
As we consider relevant only (positive energy) states around $k_F$, we will keep only these states in the expansion over ${\bf k}$-eigenstates of field $\chi({\bf r})$, and, further, only consider the BCS pairing channel in $V_{st}$.
Thus
\begin{equation}
\chi({\bf r}) = \frac{1}{\sqrt{2V}} \sum_{{\bf k}} \exp\{i {\bf k}{\bf r}\} \chi_E ({\bf k}) a_{\bf k} + \cdots ,
\label{exp}
\end{equation}
and
\begin{eqnarray}
V_{st}^{BCS} & = & \frac{2 \pi}{8 V} \sum_{{\bf k},{\bf p}} a_{{\bf k}}^\dagger a_{{\bf p}} a_{-{\bf k}}^\dagger a_{-{\bf p}}
 \nonumber \\
& \times & \frac{1}{E_{{\bf k}} E_{{\bf p}} (m + E_{{\bf k}}) (m + E_{{\bf p}}) }  \nonumber \\
& \times &\left[
  \begin{array}{cc}
   m + E_k  & k_-
  \end{array}
\right]
\left[
  \begin{array}{cc}
   0 & \frac{1}{k_+ - p_+} \\
 - \frac{1}{k_- - p_-}  & 0 \\
  \end{array}
\right]
\left[
  \begin{array}{c}
   m + E_p \\
   p_+ \\
  \end{array}
\right]
\nonumber \\
& \times & [(m + E_k) (m + E_p) + k_- p_+ ]. \nonumber \\
\label{maineq}
\end{eqnarray}
We used:
$ \int d{\bf r} \frac{1}{z} \exp\{i {\bf k}{\bf r}\} = i \frac{2 \pi}{k_+}.$
The terms with the coefficient $ k_- p_+$ give a $p$-wave channel contribution (for spinless fermions),
\begin{eqnarray}
&&\frac{k_- p_+}{|{\bf k} - {\bf p}|^2} \times \nonumber \\
&&\{(2 m + E_k + E_p )(m + E_k )(m + E_p ) - \nonumber \\
&& |p|^2 (m + E_k ) - |k|^2 (m + E_p )\}.
\end{eqnarray}
These terms give the following contribution
\begin{eqnarray}
V_{p}^{BCS} & = & m  \frac{2 \pi}{2 V} \sum_{{\bf k},{\bf p}} a_{{\bf k}}^\dagger a_{{\bf p}} a_{-{\bf k}}^\dagger a_{-{\bf p}}
\times \nonumber \\
&& \exp\{- i (\theta_k  - \theta_p )\}
  \frac{|k| |p|}{E_{{\bf k}} \cdot E_{{\bf p}} |{\bf k} - {\bf p}|^2 } ,
\label{diracchannel}
\end{eqnarray}
where we see that because of the assumed  sign of the mass, $m < 0$, we have an attractive pairing channel.
We will discuss in more detail the effective interaction and possible pairing solutions below, but in the following we will make
a few general comments. We see from Eq. (\ref{diracchannel}) that only
for non-zero mass we can have pairing. Also the chirality of the induced $p$-wave pairing can be identified. Notice the different phase factors in $V_{p}^{BCS}$ with respect to Ref. [\onlinecite{cai}]. That comes from different overall phase in eigenstates that we used in the fermion field expansion in Eq. (\ref{exp}) and the one used in Ref. [\onlinecite{cai}]. Both representations lead to a special chirality pairing function $g({\bf r})$:
\begin{equation}
\lim_{|r| \rightarrow \infty} g({\bf r}) \sim f(|{\bf r}|) \frac{|z|}{z}.
\end{equation}
The function $f(|{\bf r}|)$ depends on the details of the small ${\bf k}$ behavior of the order parameter.
 The self-consistent equation for the pairing function, $ \Delta_{\bf k}^* = 2 {\cal E}_{\bf k} \langle a_{\bf k}^\dagger a_{-{\bf k}}^\dagger \rangle $ where ${\cal E}_{\bf k}^2 = (E_{{\bf k}}- \mu )^2 + |\Delta_{\bf k}|^2 $, implied by Eq. (\ref{diracchannel}), with the assumption that  $|\Delta_k |$ is the largest around $|{\bf k}| = k_F$, give us the small ${\bf k}$ behavior,  $ \Delta_{\bf k}^* \sim   k_+ $, and thus $ g({\bf r}) \sim \frac{1}{z}$.  But again we have to take into account that the assumed direction of the external magnetic field in the Son's formalism is $\vec{B} = B \; \vec{e}_z, \; B > 0$, because the
uniform Dirac composite fermions density, $\bar{\rho}_\chi =
\vec{\nabla} \times \vec{A} = B > 0$. For the usual set-up, with $B < 0$, the analysis implies $ g({\bf r}) \sim  \frac{1}{z^*}$, i.e. Pfaffian pairing of the opposite chirality with respect to the one given by the external field.
The same conclusions i.e. attractive pairing channel with special PH Pfaffian chirality pairing hold true for $m > 0$ as can be easily checked. Thus for large enough mass we may expect that the interaction term due to the gauge field (Eq. (\ref{diracchannel}))  can lead to the PH Pfaffian-like pairing instability. But for very large $m$ the pairing interaction  is suppressed.  (For large $m$, in the scope of the HLR theories as shown in Appendix C, any pairing (Pfaffian and anti-Pfaffian) correlations that come from the current-density interaction are obstructed by a three-body interaction, and do not give a clear scenario that comes from the constrained dynamics of the system.)

A more careful examination of the Cooper channel interaction in Eq. (\ref{diracchannel}), which we may begin by angular integration in a BCS self-consistent equation, shows that the pairing interaction is extremely singular and would overcome any repulsive, short-range or Coulomb, interaction. Also the interaction in Eq. (\ref{diracchannel}) does not correspond exactly to the statistical interaction that is usually connected with the Pfaffian physics as described in Ref. [\onlinecite{gww}]. Thus we need to examine more carefully
all the terms that follow from Eq.(\ref{maineq}). We can rewrite the Cooper channel interaction in Eq.(\ref{maineq}),
taking into account both possibilities for the sign of mass,
\begin{eqnarray}
V_{st}^{BCS} & = & - \frac{m}{|m|} \frac{2 \pi}{8 V} \sum_{{\bf k},{\bf p}} a_{{\bf k}}^\dagger a_{{\bf p}} a_{-{\bf k}}^\dagger a_{-{\bf p}}
 \nonumber \\
& \times & \frac{1}{E_{{\bf k}} \cdot E_{{\bf p}} |{\bf k} - {\bf p}|^2 } \nonumber \\
 & \times & \{ - |{\bf p}|^2 ( m + E_k ) - |{\bf k}|^2 ( m + E_p ) + \nonumber \\
& & \; + \; 4 m k_- p_+ \nonumber \\
 & & \; + \;  \frac{(k_- p_+ )^2}{|{\bf p}|^2 |{\bf k}|^2} (E_k + E_p + 2 m) (E_k - m)(E_p - m)\}. \nonumber \\
\end{eqnarray}
We expect that in a self-consistent BCS equation the most important contribution will come from the region in which $ {\bf k} \approx {\bf p}$, due to the denominator in the equation above. To explore this limiting behavior, we can divide terms in curly brackets as follows,
\begin{eqnarray}
  &&  [ - |{\bf p}|^2 ( m + E_k ) - |{\bf k}|^2 ( m + E_p )  \nonumber \\
 && \; + \; 2 m (k_- p_+ + k_+ p_- ) \nonumber \\
 & &\;  +  \;\frac{(E_k + E_p + 2 m) (E_k - m)(E_p - m)}{2 |{\bf p}|^2 |{\bf k}|^2} \nonumber \\
 & &\;\; \times  ((k_- p_+ )^2 + (k_+ p_- )^2 )] \nonumber \\
 & & \; + [ 2 m (k_- p_+ - k_+ p_- ) \nonumber \\
 &  & \; + \frac{(E_k + E_p + 2 m) (E_k - m)(E_p - m)}{2 |{\bf p}|^2 |{\bf k}|^2} \nonumber \\
& & \;\; \times ((k_- p_+ )^2 - (k_+ p_- )^2 )]. \nonumber \\
 \end{eqnarray}
The first part in the square brackets is an even function of $ (\theta_p - \theta_k )$ and as $ {\bf k} \rightarrow {\bf p}$ the part is of the order of $(\theta_p - \theta_k )^2 $. The second part is leading and dominant because in the same limit  it is of the order of $ (\theta_p - \theta_k )$. The Cooper channel can be cast in the
following form
\begin{eqnarray}
V_{st}^{BCS} & = & \frac{2 \pi}{8 V }  \sum_{{\bf k},{\bf p}} a_{{\bf k}}^\dagger a_{{\bf p}} a_{-{\bf k}}^\dagger a_{-{\bf p}} \frac{1}{E_{{\bf k}} \cdot E_{{\bf p}} }
 \nonumber \\
 & \times & \{- 4 |m||{\bf k}| |{\bf p}| \frac{i \sin (\theta_p - \theta_k )}{|{\bf k} - {\bf p}|^2 } \nonumber \\
 && - \frac{m}{|m|} (E_k + E_p + 2 m) (E_k - m)(E_p - m) \nonumber \\
  && \times \frac{i \sin 2(\theta_p - \theta_k )}{|{\bf k} - {\bf p}|^2 }  \nonumber \\
&& + 4 |m||{\bf k}| |{\bf p}| \frac{(\lambda - 1 )}{|{\bf k} - {\bf p}|^2 } \nonumber \\
&& - 4 |m||{\bf k}| |{\bf p}| \frac{ \cos (\theta_p - \theta_k ) - 1}{|{\bf k} - {\bf p}|^2 } \nonumber \\
&& - \frac{m}{|m|} (E_k + E_p + 2 m) (E_k - m)(E_p - m) \nonumber \\
  && \times \frac{ \cos 2(\theta_p - \theta_k ) - 1}{|{\bf k} - {\bf p}|^2 }  \},
 \label{BSCeff}
\end{eqnarray}
where $ \lambda = \frac{|{\bf k}|^2 + |{\bf p}|^2}{2 |{\bf k}||{\bf p}|}$.
The following analysis of the effective Cooper channel in Eq. (\ref{BSCeff})  is based on considerations similar to those described in the case of classical composite fermions Pfaffian pairing in Ref. [\onlinecite{gww}].

For $ m > 0 $ and large, the Cooper channel can be approximated as,
\begin{eqnarray}
V_{st}^{BCS} & \approx &  \frac{2\pi}{8 V }  \sum_{{\bf k},{\bf p}} a_{{\bf k}}^\dagger a_{{\bf p}} a_{-{\bf k}}^\dagger a_{-{\bf p}} \frac{1}{E_{{\bf k}} \cdot E_{{\bf p}} }
 \nonumber \\
 & \times & - 2 |m| \{ \frac{i \sin (\theta_p - \theta_k )}{\lambda - \cos (\theta_p - \theta_k ) } - 1 \}.
\end{eqnarray}
Thus, as previously discussed, the implied angular momentum pairing is $\Delta_{{\bf k}}^* \sim
\langle a^\dagger_{{\bf k}} a^\dagger_{- {\bf k}} \rangle \sim e^{i \theta_{{\bf k}}}$ i.e. a PH Pfaffian-like pairing. For $ m > 0 $ and small, the Cooper channel can be approximated as,
\begin{eqnarray}
V_{st}^{BCS} & \approx &  \frac{2 \pi}{8 V }  \sum_{{\bf k},{\bf p}} a_{{\bf k}}^\dagger a_{{\bf p}} a_{-{\bf k}}^\dagger a_{-{\bf p}} \frac{1}{E_{{\bf k}} \cdot E_{{\bf p}} }
 \nonumber \\
 &\times & - \frac{m}{|m|} (E_k + E_p + 2 m) (E_k - m)(E_p - m) \nonumber \\
  && \times \frac{ \exp i 2(\theta_p - \theta_k ) - 1}{|{\bf k} - {\bf p}|^2 }. \label{effAPF}
\end{eqnarray}
We find, by doing the angular integration first in the implied BCS self-consistent equation, that the pairing
$\Delta_{{\bf k}}^* \sim e^{i \theta_{{\bf k}}}$ is suppressed, and that $\Delta_{{\bf k}}^* \sim e^{i 3 \theta_{{\bf k}}}$ is the dominant pairing. The pairing in the same direction of PH Pfaffian, with angular momentum equal to 3, can be identified as an anti-Pfaffian instability. For $ m < 0$ and $|m|$ small, the sign of the effective Cooper channel in Eq. (\ref{effAPF}) is switched. This changes the chirality of the implied pairing, and we find that now
$\Delta_{{\bf k}}^* \sim e^{- i \theta_{{\bf k}}}$ is the dominant pairing, which we can identify with a Pfaffian
instability. For $ m < 0$ and $|m|$ large, the effective channel is
\begin{eqnarray}
V_{st}^{BCS} & \approx &  \frac{2 \pi}{8 V }  \sum_{{\bf k},{\bf p}} a_{{\bf k}}^\dagger a_{{\bf p}} a_{-{\bf k}}^\dagger a_{-{\bf p}} \frac{1}{E_{{\bf k}} \cdot E_{{\bf p}} }
 \nonumber \\
 & \times & - 2 |m| \{ \frac{i \sin (\theta_p - \theta_k )}{\lambda - \cos (\theta_p - \theta_k ) } \nonumber \\
  && - \lambda
 \frac{ (\exp i 2 (\theta_p - \theta_k ) - 1)}{\lambda - \cos (\theta_p - \theta_k ) } \},
\end{eqnarray}
and we recover again a PH Pfaffian instability, $\Delta_{{\bf k}}^* \sim e^{ i \theta_{{\bf k}}}$.

The picture that emerges from our analysis is very simple: for large $|m|$ irrespective of the sign of mass, we may expect PH Pfaffian-like state, but for small $|m|$, depending on its sign, we have anti-Pfaffian,
for $m > 0$, and Pfaffian state for $m < 0$. Due to considerable numerical support for anti-Pfaffian under LL mixing in the SLL [\onlinecite{zal}], we may identify the case with the positive mass to the one of the SLL. Furthermore, the identification of Pfaffian and anti-Pfaffian for opposite sign of $m$ i.e. particle-hole symmetry breaking that is not large is consistent with the numerics (in the SLL) [\onlinecite{dh}].

But we should be aware of the absence of pairing in the LLL, and that our analysis based on the gauge field description only, is not sufficient for the explanation of the physics in the LLL. We need to include Coulomb repulsive interactions among electrons. This inclusion in the Chern-Simons theories, especially the DCF theory is not an easy task, because a part of the influence of the interactions is built in the gauge dynamics. We may try  to include a bare Coulomb interaction with densities that correspond to those of the Dirac quasiparticles (of the theory in the Eq. (\ref{impL})) as a consequence of Eqs. (\ref{edj}) and (\ref{fdj}). The singular behavior of the Coulomb law can suppress any pairing correlations that follow from gauge field description and constraints. Thus we need to include the interactions in a way that reflects the physics of a fixed LL
to explain the dichotomy of the physics in the LLL and SLL, i.e. Fermi-liquid-like state, and topological paired state, respectively i.e. to include more intra Landau level physics in the DCF theory. The way to achieve that is to include a term that represents the interaction of the effective dipoles of the Read's construction with an electric field as discussed in Refs. [\onlinecite{pr,ns}].  As explained in Ref.
[\onlinecite{ns}] the inclusion of this physics amounts to a change in the expression of the Coulomb interaction of the form,
\begin{eqnarray}
\frac{\alpha}{|{\bf q}|} \rightarrow \frac{\alpha}{|{\bf q}| + \frac{m^*}{2 B} \alpha
 |{\bf q}|^2} , \label{CoulombMod}
\end{eqnarray}
where we assumed a static case i.e. no external fields except for the uniform, constant magnetic field $B$, and $m^*$ represents an effective parameter (mass) in the long distance limit. In the following we briefly recapitulate how we can reach the modified interaction in (\ref{CoulombMod}). First we note that in a functional formulation we can introduce a scalar field, $\phi$, that decouples the Coulomb term in the inverse space as
\begin{eqnarray}
\delta {\cal L}_c  =  - \frac{2\pi \alpha}{|{\bf q}|} \delta \rho (- {\bf q}) \delta \rho ({\bf q}) \rightarrow  \nonumber \\
\phi (- {\bf q}) \delta \rho ({\bf q}) + \frac{|{\bf q}|}{2\pi \alpha}  \phi(- {\bf q}) \phi ({\bf q}).
\end{eqnarray}
The scalar represents a potential that a particle experiences due to other particles. On the other hand, Galilean invariance allows an extra term in the kinetic part of the DCF theory [\onlinecite{pr}],
\begin{equation}
\delta {\cal L}_u = i u_i \chi^\dagger \partial_i \chi, \label{gal_term}
\end{equation}
where $u_i $ is the local drift velocity, $ u_i = \epsilon_{ij} \frac{\partial_i \phi}{B} $, $i = x, y$. This term
represents an interaction between the (local) electric field and dipoles of the composite fermion quasiparticles, which are propotional to quasiparticle momenta [\onlinecite{pr}]. If we  introduce a mass parameter, $ m^* $, to relate the
momenta of quasiparticles to their (local) velocity ${\bf u}$, we may represent (\ref{gal_term}), in the inverse space, as
\begin{equation}
\delta {\cal L}_u = \frac{ m^* |{\bf u}|^2 \bar{\rho}}{2} =  m^* \frac{\bar{\rho}}{2 B^2} |{\bf q}|^2 \phi(- {\bf q}) \phi ({\bf q}), \label{gal_term_exp2}
\end{equation}
where $ \bar{\rho} = \frac{1}{2\pi l_B^2} $ with $ l_B = 1/k_F $, the magnetic length, is the density of the system.
Integrating field $\phi$ in the functional representation of the theory, with $\delta {\cal L}_u $ and $\delta {\cal L}_c $ included, we reach (\ref{CoulombMod}).
Thus the BCS channel in Eq. (\ref{BSCeff}) with a modified Coulomb interaction in Eq. (\ref{CoulombMod}) may represent a good starting point for the investigation of the pairing instabilities at half-filling in the presence of the PH symmetry breaking mass $m$.

The role of the modified Coulomb interaction is crucial for the existence of paired states. For $m^*$ finite, we have to deal with a singular repulsive interaction at this level of approximation, which will preclude any pairing as is the case in the LLL.
For $m^*$ infinite, the effects of the interaction will be obliterated, and we  will have the pairing scenario as is the case in the SLL. Moreover, in this case, for $|m|$ (LL mixing) large we may expect the PH Pfaffian-like state, which is stabilized with $|m|$ in a uniform system and a consequence of the constrained - gauge field description. Nevertheless we should note that PH Pfaffian effective (attractive) interaction scales as $ \sim \frac{k_F}{|m|}$ (with respect to those of Pfaffian and anti-Pfaffian for small $|m|$), and thus it is suppressed in
magnitude with large $|m|$.

We may ask ourselves what is the physical meaning of the $m^*$ infinite mass limit in the SLL. In this case the local
drift velocity should go to zero and thus the potential that other particles make for a given one is flat i.e. the
correlation hole does not exist and particles are free to pair. That this indeed may be the case in the SLL we have indications from numerical experiments that find larger size of hole excitations in the SLL than in the LLL in the
FQHE regime  at filling factors 1/3 and 7/3 [\onlinecite{sjzp}].

The numerical solutions of the BCS self-consistent equation:
\begin{equation}
\Delta_{\bf p}^* = - \sum_{\bf k}  V_{{\bf k} {\bf p}} \frac{\Delta^{*}_{{\bf k}}}{{\cal E}_{\bf k}},
\label{sceq}
\end{equation}
for channels $ l = 1,3,-1,$ with $\Delta_{\bf k}^* = |\Delta_{\bf k} | e^{i l \theta_{\bf k}}$, are described  in Fig. 1. Details concerning Eq. (\ref{sceq}) and its solutions can be found in Appendix D.
\begin{figure*}
\centering
\includegraphics[width=5.5in]{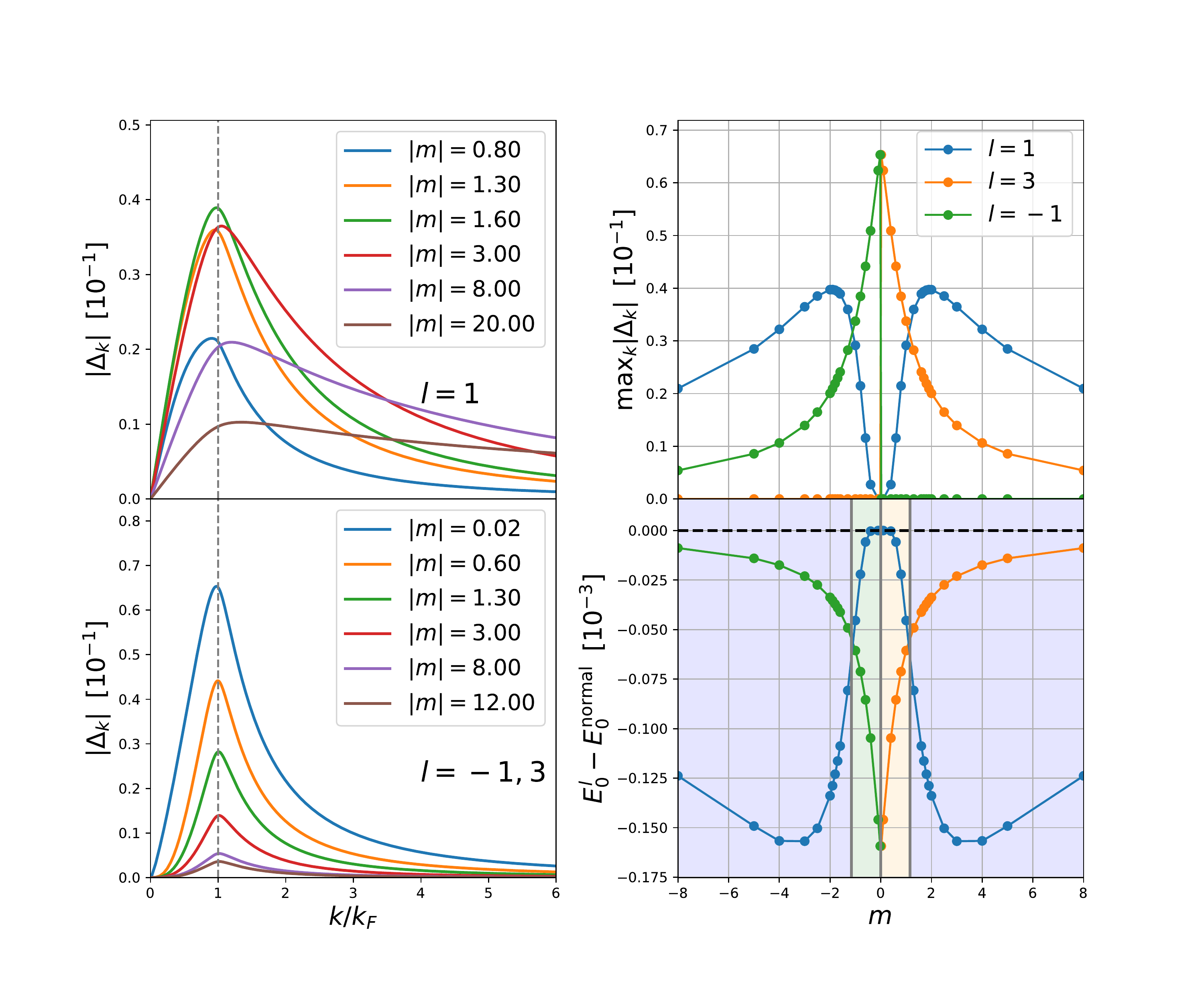}
\caption{The solutions of the self-consistent BCS problem. Left column: radial direction $k$-dependent pairing amplitude for various values of $m$. Channel $l=1$ solution only depends on $|m|$, while $l=3$ and $l=-1$ channel solutions are symmetric with the sign-flip of $m$ (see Appendix \ref{app:V_symmetry}). Upper right panel: dependence of the maximum of the pairing amplitude on $m$ (always found at the Fermi level $k_F$). Lower right panel: total energy of the different superconducting solutions compared to the normal state energy. Gray vertical lines denote the transition between different $l$ channels. Color in the background corresponds to the energetically favorable channel at the given $m$.}
\end{figure*}
The parameter $m$ in Fig. 1 is measured in units of energy, $(\hbar v_F ) k_F $, and this dimensionless quantity along $x$
axis on the right-hand side  of Fig. 1, can be described in the following way. First we rewrite the quantity with the Fermi velocity and explicit physical constants,
\begin{equation}
m \rightarrow \frac{m_D v_F^2}{\hbar v_F k_F} = \frac{m_D v_F}{\hbar } l_B = \sqrt{\frac{c}{\hbar e}} v_F \frac{m_D}{\sqrt{B}},
\end{equation}
where $m_D$ is the mass of DCFs, $l_B = \sqrt{\frac{\hbar c}{e B}}$ is the magnetic length, and $k_F = \frac{1}{l_B}$.
On the other hand the coefficient of the LL mixing is the ratio of the characteristic Coulomb energy and cyclotron energy,
\begin{equation}
\frac{V_c}{\hbar \omega_c} = \frac{e^2}{l_B} \frac{m_e c}{e B} = \frac{e \sqrt{e c}}{\hbar \sqrt{\hbar}} \frac{m_e}{\sqrt{B}},
\end{equation}
where $m_e$ is the mass of electron. Thus the plotted (dimensionless) parameter $m$ may be identified with LL mixing if $m_D = \frac{e^2}{\hbar v_F } m_e $ i.e. the mass of DCF is the electron mass multiplied with  a ``fine-structure constant" of DCFs, $ \frac{e^2}{\hbar v_F}$, characterizing the relative strength of the Coulomb interaction. The identification seems plausible, although we do not have an explicit proof; we expect that the prediction of the phase
diagram that follows from the theory, up to physical constants, depends solely on the unique parameter of the system, $ l_B = \frac{1}{k_F}$, and thus the dimensionless parameter in Fig. 1 should represent LL mixing.

The LL mixing in experiments is of order 1, although it can be large as 4-8 [\onlinecite{smd,pfj}], and with the above identification we may expect anti-Pfaffian $(l = 3)$ to be dominant instability in the SLL from the phase diagram in Fig. 1, though the critical $m$ (for the
transition into the PH Pfaffian-like state $(l = 1)$), may be estimated to be, $m_c = 1.2 $, and thus the role and
possibility for the development of a PH Pfaffian-like state, at sufficiently large LL mixing in a uniform system
should not be underestimated or excluded.

We have confidence in our predictions, because global features of the phase diagram in Fig, 1 are in agreement with
numerical experiments in SLL: (a) At $ m = 0$ a Schroedinger cat superposition of Pfaffian and anti-Pfaffian is present as in Ref. [\onlinecite{dh}], and depending on the sign of the mass for $m = 0$ we have Pfaffian and anti-Pfaffian, (b) PH Pfaffian is continuously connected to the excited composite FL state at $m = 0$ in an agreement with Ref. [\onlinecite{mish}].

\section{Conclusions}

In this work we discussed the effective Cooper pairing channel of the system at
half-filling in the scope of the DCF theory with a mass term. The mass term of the DCF theory represents a term that breaks the particle-hole symmetry of electrons confined in a LL and represents a LL mixing.
Solely on the basis of
gauge field description, we find for small Dirac mass anti-Pfaffian or Pfaffian instability depending on the sign of mass, consistent with numerical investigations of the SLL [\onlinecite{dh}], while for large mass (LL mixing),
beyond the reach of numerical experiments,
irrespective of mass sign, PH Pfaffian-like instability.

\begin{acknowledgments}
We would like to thank Vladimir Juri\v ci\'c and Michael Peterson for discussions. We also thank Nordita Institute for hospitality during the final stages of this work.
 This research was  supported by
the Ministry of Education, Science, and Technological
Development of the Republic of Serbia under project
ON171017.
\end{acknowledgments}

\appendix
\section{The polarization tensor}
The derivation of the polarization tensor $\Pi({\bf k}, k_0)$ in the massive case, in the hydrodynamic approximation, can be done following and generalizing the procedure described in the massless case in Ref. [\onlinecite{mir}]. We start from the fermion propagator in Eq. (\ref{fer_prop_def}) as described in Ref. [\onlinecite{cm}],
\begin{eqnarray}
&&i S_F (x, y) = \nonumber \\
&& \theta(x^0 - y^0) \nonumber \\
&& \times \int \frac{d^2 p}{(2\pi)^2} \frac{1}{2 p_0}
( \gamma p + m) \theta(p^0 - \mu) \exp\{-i p (x - y)\} \nonumber \\
&& - \theta(y^0 - x^0) \nonumber \\
&& \times \int \frac{d^2 p}{(2\pi)^2} \frac{1}{2 p_0}
( \gamma p + m) \theta(\mu - p_0 ) \exp\{-i p (x - y)\} \nonumber \\
&& - \theta(y^0 - x^0) \int \frac{d^2 p}{(2\pi)^2} \frac{1}{2 p_0}
( \gamma p - m)  \exp\{i p (x - y)\},  \nonumber \\  \label{propagator_described}
\end{eqnarray}
where $ p_0 = \sqrt{{\bf p}^2 + m^2}, p x = p_0 x^0 - {\bf p} {\bf x}$, and
$ \gamma p = \gamma^0 p_0 - {\bf \gamma} {\bf p}$. Using the theta function
representation,
\begin{equation}
\theta(x^0 - y^0) = - \int \frac{d \omega}{2\pi i} \frac{\exp\{-i \omega (x^0 - y^0)\}}{\omega + i \eta}
\end{equation}
we arrive at
\begin{eqnarray}
&& i S_F (x - y) =  \nonumber \\
&& \int \frac{d \omega}{2\pi} \int \frac{d^2 p}{(2\pi)^2} \times \nonumber \\
&& \exp\{-i \omega (x^0 - y^0) + i {\bf p} ({\bf x} - {\bf y})\} \times \nonumber \\
&& [ i \gamma^0 \frac{1}{\omega - p_0 + i \eta (\theta(p_0 - \mu) - \theta(\mu - p_0))} \Omega_{\bf p}^{-} + \nonumber \\
&& i \gamma^0 \frac{1}{\omega + p_0 - i \eta } \Omega_{\bf p}^{+}] \label{prodes0}
\end{eqnarray}
where
\begin{equation}
\Omega_{\bf p}^{-} = \frac{1}{2} (1 + \frac{\gamma^0 m}{p_0} - \frac{\gamma^0 {\bf \gamma} {\bf p}}{p_0}),
\label{prodes1}
\end{equation}
and
\begin{equation}
\Omega_{\bf p}^{+} = \frac{1}{2} (1 - \frac{\gamma^0 m}{p_0} + \frac{\gamma^0 {\bf \gamma} {\bf p}}{p_0}).
\label{prodes2}
\end{equation}
To get the form of the fermion operator in Ref. [\onlinecite{mir}] we can shift the frequency variable $\omega$ as $ \omega \rightarrow \omega + \mu $.
To find
\begin{equation}
\Pi^{\mu \nu} = - i tr[ \gamma^\mu S_F (x, y)\gamma^\nu S_F (y, x)], \label{defPi2}
\end{equation}
we need to generalize the trace calculations, frequency, and momentum integrals. The main approximation in the momentum integrals for the external momentum ${\bf k}$, $|{\bf k}| \ll k_F, \mu $, and internal momentum ${\bf q}$, $|{\bf q}| \sim k_F $ (constrained by the Fermi statistics inside integrals for $|{\bf k}| \ll k_F $) is
\begin{eqnarray}
&& \theta(\sqrt{|{\bf q} + {\bf k}|^2 + m^2} - \mu) \approx \nonumber \\
&& \theta(\sqrt{|{\bf q}|^2 + m^2} - \mu + |{\bf k}| \frac{|{\bf q}|}{\sqrt{|{\bf q}|^2 + m^2} } \cos \phi) \approx \nonumber \\
&& \theta(\sqrt{|{\bf q}|^2 + m^2} - \mu) + \nonumber \\
&&  |{\bf k}| \frac{k_F}{\sqrt{k_F^2 + m^2}} \delta(\sqrt{|{\bf q}|^2 + m^2} - \mu) \cos \phi,
\end{eqnarray}
where $k_F = \sqrt{\mu^2 - m^2}$, and $\phi$ is the angle between vectors ${\bf k}$ and ${\bf q}$. Here an important difference with respect to the massless case is the appearance of the factor $\frac{k_F}{\sqrt{k_F^2 + m^2}}$.
A detailed analysis leads to a conclusion that to get $\Pi^{\mu \nu}$ in the massive case we have to rescale external momenta ${\bf k}$ in the massless case with factor $ \alpha_F = \frac{k_F}{\sqrt{k_F^2 + m^2}}$ i.e. $ {\bf k} \rightarrow \alpha_F {\bf k}$. In particular,
\begin{equation}
\Pi^{00}(k_0, {\bf k}) = \tilde{\Pi}^{00}(k_0, \alpha_F {\bf k}), \label{1}
\end{equation}
\begin{eqnarray}
&& \Pi^{0i}(k_0, {\bf k}) = \alpha_F \tilde{\Pi}^{0i}(k_0, \alpha_F {\bf k}) +
\nonumber \\
&& \epsilon_{ji} ( - \frac{i k_j m}{4 \pi} - i \frac{k_j m}{2 \mu^2} \tilde{\Pi}^{00}(k_0, \alpha_F {\bf k}))
\label{antisym}
\end{eqnarray}
\begin{equation}
\Pi^{ij}(k_0, {\bf k}) = \alpha_F^2 \tilde{\Pi}^{ij}(k_0, \alpha_F {\bf k}),\label{3}
\end{equation}
where by $\tilde{\Pi}^{\mu \nu}; \mu, \nu = 0, i, j$ we denoted the components of the polarization tensor in the massless case, and, furthermore, we can see the
antisymmetric (Hall conductance) contribution in $\Pi^{0i}$, due to the presence of the mass term.

The components of the polarization tensor in the massless case, $\tilde{\Pi}^{\mu \nu}(k_0, {\bf k})$, can be found in Ref. [\onlinecite{mir}], and they are
\begin{eqnarray}
\tilde{\Pi}^{00}(k_0, {\bf k}) & = & \Pi_l(k_0, {\bf k}), \nonumber \\
\tilde{\Pi}^{0i}(k_0, {\bf k}) & = & k_0 \frac{k^i}{|{\bf k}|^2} \Pi_l(k_0, {\bf k}), \nonumber \\
\tilde{\Pi}^{ij}(k_0, {\bf k}) & = & (\delta^{ij} -  \frac{k^i k^j}{|{\bf k}|^2}) \Pi_t(k_0, {\bf k}) + \nonumber \\
&& \frac{k^i k^j}{|{\bf k}|^2} \frac{k_0^2 }{|{\bf k}|^2} \Pi_l(k_0, {\bf k}), \label{4}
\end{eqnarray}
where
\begin{eqnarray}
&& \Pi_l(k_0, {\bf k}) = \nonumber \\
&& \frac{\mu}{2 \pi} (\theta(k^2)\sqrt{\frac{k_0^2}{k^2}} - 1 - i \theta(- k^2)\sqrt{\frac{k_0^2}{- k^2}}),\label{L}
\end{eqnarray}
and
\begin{equation}
\Pi_t(k_0, {\bf k}) = \frac{\mu}{2 \pi} - \frac{k^2}{|{\bf k}|^2} \Pi_l(k_0, {\bf k}), \label{5}
\end{equation}
and $ k^2 = k_0^2 - |{\bf k}|^2 $.

The antisymmetric contribution in Eq. (\ref{antisym}) is expected from the Berry curvature contributions in the scope of the relativistic quantum mechanics [\onlinecite{pot}], and here the Hall conductance can be recovered to be, $\sigma_H = -  \frac{1}{4 \pi} \frac{m}{\sqrt{k_F^2 + m^2}}$.

It is important to comment that due to the infinite Dirac sea, we have divergent contributions to the polarization tensor (when doing the calculation according to the definition). As discussed in the main text the DCF theory as defined in Eq. (\ref{mL}) requires dimensional regularization in order to recover finite $\Pi^{\mu \nu}$. We used a version of the DCF theory, given in Eq. (\ref{impL}) and associated Pauli-Villars regularization to recover $\Pi^{\mu \nu}$.
\section{The propagator of the gauge field}
To find $\bar{\cal D}_{\mu \nu}^{-1}$ defined in Eq.(\ref{invDbar}) we need to switch from Minkowski to Euclidean space - time. According to the definition of $\bar{\cal D}_{\mu \nu}^{-1}$ we need only to take into account the change in the fermion propagator, which amounts to taking $i \omega$ instead of $\omega$ at $T({\rm temperature}) = 0$ in the Fourier transform of the fermion propagator described in Eqs. (\ref{prodes0}-\ref{prodes2}). Thus we have to repeat steps that we took to calculate $ \Pi^{\mu \nu}$ taking into account this change. The components of $\bar{\cal D}_{\mu \nu}^{-1}$ are formally equal to the expressions in Eqs. (\ref{1}-\ref{4},\ref{5}) i.e. $\bar{\cal D}_{\mu \nu}^{-1}(k_0, {\bf k}) = \Pi_{\mu \nu}(k_0, {\bf k})$, with $\Pi_l(k_0, {\bf k})$ equal to
\begin{eqnarray}
&& \hat{\Pi}_l (k_0, {\bf k}) = \nonumber \\
&& \frac{\mu}{2 \pi} (- 1 + \frac{1}{\sqrt{1 - \frac{|{\bf k}|^2}{k_0^2}}}),
\end{eqnarray}
where $k_0$ is purely imaginary.

In the transverse gauge, $\vec{\nabla} {\bf a} = 0$, and if we denote by $\alpha = \frac{2 \pi e^2}{\epsilon_r}$, the Coulomb coupling constant, the inverse of the gauge field propagator is
\begin{equation}
{\cal D}^{-1}(k_0 ,{\bf k}) = \left[
  \begin{array}{cc}
   \hat{\Pi}_{00}  &  \hat{\Pi}_{0T} \\
   \hat{\Pi}_{0T} &  \hat{\Pi}_{TT}\\
  \end{array} \label{b2}
\right],
\end{equation}
where
\begin{equation}
\hat{\Pi}_{00} =  \hat{\Pi}_l (k_0, \alpha_F {\bf k}), \label{b3}
\end{equation}
\begin{equation}
\hat{\Pi}_{0T} = - \frac{1}{4 \pi} \frac{m}{\mu} |{\bf k}| -  \frac{m}{2 \mu^2} |{\bf k}|  \hat{\Pi}_l (k_0, \alpha_F {\bf k}), \label{b4}
\end{equation}
\begin{equation}
\hat{\Pi}_{TT} =  \frac{k_F^2}{2 \pi \mu} - \frac{k_0^2 - \alpha_F^2 |{\bf k}|^2}{|{\bf k}|^2} \hat{\Pi}_l (k_0, \alpha_F {\bf k})
+ \alpha |{\bf k}| \frac{1}{(4 \pi)^2} . \label{b5}
\end{equation}
Here we defined the transverse component of the gauge field to be $a_T = i {\bf k} \times {\bf a}({\bf k})$ and it
is understood that $k_0$ is purely imaginary.

\section{Classical HLR fermions at half-filling and pairing instabilities}
It is interesting to probe the large $|m|$ limit of the Lagrangian given by Eq.(\ref{impL}). In this limit and after redefinitions $a_\mu \rightarrow a_\mu + \frac{m}{|m|} A_\mu$, the Lagrangian becomes,
\begin{eqnarray}
{\cal L}_{\text{ccf}} & = &  \frac{m}{|m|} \frac{1}{8 \pi} a da + \psi^\dagger (i \partial_0 + a_0 +  \frac{m}{|m|} A_0 ) \psi  \nonumber \\
 && - \sum_{i=x,y} \frac{1}{2 |m|} \psi^\dagger (i \partial_i + a_i +  \frac{m}{|m|} A_i )^2 \psi  \nonumber \\
 && + \frac{(1 - \frac{m}{|m|}) }{2} \frac{1}{4 \pi} A dA .  \label{clasL}
\end{eqnarray}
If $m > 0$ we have the usual Lagrangian of HLR (up to a Coulomb interaction term that we omitted), which, based on the mean field approximation, leads to the description of composite fermion liquid (CFL). On the other hand, for $ m < 0 $, we have exactly the Lagrangian of Ref. \onlinecite{bmf}, which in the same approximation  describes anti-CFL i.e. Fermi liquid of composite holes.

The HLR theory ($m > 0$ case) is described by the following kinetic term of the Hamiltonian density,
\begin{equation}
{\cal K} = \frac{1}{2 |m|} (- i {\bf \nabla} + {\bf c}) \psi^\dagger
(i {\bf \nabla} + {\bf c}) \psi ,
\label{kinetic}
\end{equation}
where $\vec{c}$ represents the deviation from the uniform magnetic field configuration of the CS gauge field  $a_\mu$ : $c_\mu = a_\mu + A_\mu$ such that,
\begin{equation}
- \delta \rho_\psi = \frac{1}{4 \pi} {\bf \nabla} \times {\bf c}.
\label{hlreq}
\end{equation}
In this non-relativistic case the effective statistical density-current interaction is given by
\begin{eqnarray}
\tilde{V}_{\text{st}} & = & - \frac{1}{2 |m|} {\bf c} [\psi^\dagger (i {\bf \nabla} \psi ) - (i {\bf \nabla} \psi^\dagger ) \psi ] \nonumber \\
& \equiv & \frac{{\bf c} {\bf j}}{2 |m|}
\end{eqnarray}
Using Eq. (\ref{hlreq}), in parallel to Eqs. (\ref{fdj}) and (\ref{a}) in the relativistic case, we can express the interaction as
\begin{eqnarray}
\tilde{V}_{\text{st}} & = & \frac{1}{2 |m|} \int d{\bf r}' [ \frac{y - y'}{|{\bf r} - {\bf r}'|^2} j_x   \nonumber \\
&& - \frac{x - x'}{|{\bf r} - {\bf r}'|^2} j_y ] \delta \rho({\bf r}').
\end{eqnarray}
If we introduce momentum space states,
\begin{equation}
\psi({\bf r})= \frac{1}{\sqrt{V}} \sum_{{\bf k}} \exp\{ i {\bf k} {\bf r} \} \; c_{{\bf k}},
\end{equation}
\begin{eqnarray}
 {\bf j}({\bf k})& = & \frac{1}{\sqrt{V}} \int d{\bf r}\exp\{- i {\bf k} {\bf r} \} \; {\bf j}({\bf r})   \nonumber \\
& = & \frac{1}{\sqrt{V}} \sum_{{\bf q}} (2 {\bf q} - {\bf k})\; c^\dagger_{{\bf q}} \; c_{{\bf k}+{\bf q}}.
\end{eqnarray}
Thus
\begin{eqnarray}
&& \int d{\bf r} \tilde{V}_{\text{st}} ({\bf r})   \nonumber \\
&& = \frac{i \; V}{|m|} \sum_{{\bf q}, {\bf p}, {\bf l}}
\frac{{\bf q}\times {\bf p}}{|{\bf q}|^2}\; c^\dagger_{{\bf p}} \; c_{{\bf p}-{\bf q}}\; c^\dagger_{{\bf l}} \; c_{{\bf l}+{\bf q}}
\end{eqnarray}
The BCS channel we get by taking $ {\bf l} = - {\bf p}$. If we let ${\bf p} \rightarrow {\bf k}$ and ${\bf q} \rightarrow {\bf k} - {\bf p}$ we have
\begin{eqnarray}
&& \sum_{{\bf k}} \tilde{V}_{\text{st}} ({\bf k})   \nonumber \\
&& = \frac{- i \; V}{|m|} \sum_{{\bf k}, {\bf p}}
\frac{{\bf p}\times {\bf k}}{|{\bf p} - {\bf k}|^2}\; c^\dagger_{{\bf k}} \; c_{{\bf p}}\; c^\dagger_{-{\bf k}} \; c_{-{\bf p}}
\label{hlrchannel}
\end{eqnarray}
Now we should note that
\begin{equation}
{\bf p}\times {\bf k} = \frac{p_{+} k_{-} - p_{-} k_{+}}{- 2 i}
\label{formula}
\end{equation}
Direct comparison of Eq. (\ref{hlrchannel}) with Eq. (\ref{diracchannel}) shows that a $p$-wave of opposite chirality with respect to the one of PH Pfaffian, i.e. a Pfaffian  $p$-wave, is the statistical interaction implied BCS pairing instability of classical HLR CFs.

The diamagnetic term in Eq. (\ref{kinetic}) i.e. the term $\sim {\bf c}^2 \; \psi^\dagger \psi$ makes an interesting three body interaction in the real space:
\begin{eqnarray}
&&\int d{\bf r}_3 \tilde{V}_{\text{st}}({\bf r}_3) \sim \nonumber \\
 && \int d{\bf r}_1 \int d{\bf r}_2 \int d{\bf r}_3  \frac{{\bf r}_1 - {\bf r}_3}{|{\bf r}_1 - {\bf r}_3|^2}   \frac{{\bf r}_2 - {\bf r}_3}{|{\bf r}_2 - {\bf r}_3|^2} \times \nonumber \\
 && \delta \rho({\bf r}_1) \delta \rho({\bf r}_2) \delta \rho({\bf r}_3),
\end{eqnarray}
which sign is fluctuating and this interaction represents a disordering factor.

We can easily repeat the analysis in the anti-CFL case and find that the current-density statistical interaction favors opposite chirality  pairing with respect to Pfaffian but of composite holes. This special pairing state of composite holes can be identified with anti-Pfaffian [\onlinecite{bmf}]. But again the additional, fluctuating sign 3 body interaction, next to the attractive channel exists.

\section{BCS self-consistent problem and its solutions}

We start with the relevant parts of BCS mean field theory and follow the notation  of Ref. [\onlinecite{rg}]. The
effective Hamiltonian is
\begin{equation}
K_{\rm eff} = \sum_{\bf k} \{ \xi_k c_{\bf k}^\dagger c_{\bf k} + \frac{1}{2} (\Delta^* c_{- {\bf k}} c_{\bf k} + \Delta c_{\bf k}^\dagger  c_{-{\bf k}}^\dagger) \},
\end{equation}
and in our case $\xi_k = E_k - \mu $, with $ E_k = \sqrt{|{\bf k}|^2 + m^2}$. The Bogoliubov transformation is
\begin{equation}
\alpha_{\bf k} = u_{\bf k} c_{\bf k} - v_{\bf k} c_{-{\bf k}}^{\dagger},
\end{equation}
with
\begin{eqnarray}
\frac{v_{\bf k}}{u_{\bf k}} & = & \frac{- ({\cal E}_k - \xi_k )}{\Delta_{\bf k}^{*}} , \nonumber \\
|u_{\bf k}|^2 & = & \frac{1}{2} (1 + \frac{\xi_k }{{\cal E}_k}), \nonumber \\
|v_{\bf k}|^2 & = & \frac{1}{2} (1 - \frac{\xi_k }{{\cal E}_k}),
\end{eqnarray}
and ${\cal E}_k = \sqrt{\xi^2_k + |\Delta_{\bf k}|^2}$.

On the other hand if we start with a Cooper channel interaction and do the BCS mean field decomposition with
$ b_{\bf k}^\dagger = c_{\bf k}^\dagger c_{-{\bf k}}^\dagger $
\begin{eqnarray}
&& \sum_{{\bf k},{\bf p}} V_{{\bf k} {\bf p}} \; b_{\bf k}^\dagger \; b_{{\bf p}} = \sum_{{\bf k},{\bf p}} V_{{\bf k} {\bf p}} <b_{\bf k}^\dagger> b_{\bf p}  \nonumber \\
&& + \sum_{{\bf k},{\bf p}} V_{{\bf k} {\bf p}} b_{\bf k}^\dagger <b_{\bf p}>
- \sum_{{\bf k},{\bf p}} V_{{\bf k} {\bf p}} <b_{\bf k}^\dagger> <b_{\bf p}>, \nonumber \\
\end{eqnarray}
and specify $u_{- \bf k} = u_{\bf k} = u_{\bf k}^*$ and $v_{-{\bf k}} = - v_{{\bf k}}$,
then
\begin{eqnarray}
&&\frac{\Delta^*_{\bf p}}{2}  =  \sum_{\bf k} V_{{\bf k} {\bf p}} <c_{\bf k}^\dagger c_{-{\bf k}}^\dagger> \nonumber \\
&& =  \sum_{\bf k} V_{{\bf k} {\bf p}} <(u_{\bf k} \alpha_{\bf k}^\dagger + v_{\bf k}^* \alpha_{-{\bf k}}) (- v_{\bf k}^* \alpha_{{\bf k}} + u_{\bf k} \alpha_{-{\bf k}}^\dagger)>, \nonumber \\
\end{eqnarray}
i.e.
\begin{equation}
\frac{\Delta^*_{\bf p}}{2} = \sum_{\bf k} V_{{\bf k} {\bf p}} v_k^* u_k = \sum_{\bf k} V_{{\bf k} {\bf p}} (-) \frac{\Delta_{\bf k}^*}{2 \; {\cal E}_k}, \label{apsc}
\end{equation}
and thus  Eq.(\ref{sceq}) in the main text.

\subsection{BCS equation in polar coordinates $\mathbf{k}\rightarrow(k,\theta_k)$}

We simplify the expression in Eq.(\ref{BSCeff}) to obtain
\begin{eqnarray}
V_{{\bf k}{\bf p}} & = & \frac{2 \pi}{8 V } \frac{1}{E_k \cdot E_p } \times
 \nonumber \\
 & \times & \Bigg[- 4 |m|kp \frac{i \sin (\theta_p - \theta_k )}{|{\bf k} - {\bf p}|^2 } \nonumber \\
 && \;\;- \frac{m}{|m|} (E_k + E_p + 2 m) (E_k - m)(E_p - m) \nonumber \\
  && \;\;\;\;\;\;\times \frac{ \exp\{ i 2(\theta_p - \theta_k )\} - 1}{|{\bf k} - {\bf p}|^2 } \Bigg].
 \label{Vkp}
\end{eqnarray}
For a fixed angular momentum channel, $ \Delta_{\bf k}^* = |\Delta_{\bf k}| e^{i l \theta_k}$, we do first the
integration over the angular variable, $\theta_k - \theta_p $, in Eq. (\ref{sceq}) (or Eq. (\ref{apsc}) and after the change from sum to integral:  $ \sum_{\bf k} \rightarrow \frac{V}{(2 \pi)^2} \int d{\bf k}$).
We use
\begin{equation}
I_m = \int_0^{2 \pi} d\theta \frac{\sin m \theta \sin \theta}{\lambda - \cos \theta} = 2 \pi (\lambda - \sqrt{\lambda^2 - 1})^m , \label{formule}
\end{equation}
$ m = 1, 2, 3$, with $\lambda = \frac{k^2 + p^2}{2 k p}$, to get
\begin{equation}
V_{kp}^l = \frac{1}{2 \pi} \int_0^{2 \pi} d(\theta_k - \theta_p )  e^{i l (\theta_k - \theta_p )} V_{{\bf k}{\bf p}},
\label{DefV}
\end{equation}
for $l = 1, 3, -1$.

In particular, for $l = 1$ in (\ref{DefV}), we use (\ref{formule}) to express the following integral,
\begin{equation}
\int_0^{2 \pi} d(\theta_k - \theta_p )  e^{i (\theta_k - \theta_p )} \frac{- i \sin(\theta_k - \theta_p )}{\lambda - \cos(\theta_k - \theta_p )} = I_1
\end{equation}
and
\begin{eqnarray}
&& \int_0^{2 \pi} d(\theta_k - \theta_p )  e^{i (\theta_k - \theta_p )} \frac{e^{-i 2 (\theta_k - \theta_p )} - 1}{\lambda - \cos(\theta_k - \theta_p )} = \nonumber \\
&& \int_0^{2 \pi} d(\theta_k - \theta_p )   \frac{e^{-i  (\theta_k - \theta_p )} - e^{ i  (\theta_k - \theta_p )}}{\lambda - \cos(\theta_k - \theta_p )} = 0, \nonumber \\
\end{eqnarray}
for $l = 3$ in (\ref{DefV}), we have
\begin{equation}
\int_0^{2 \pi} d(\theta_k - \theta_p )  e^{i 3 (\theta_k - \theta_p )} \frac{- i \sin(\theta_k - \theta_p )}{\lambda - \cos(\theta_k - \theta_p )} = I_3
\end{equation}
and
\begin{eqnarray}
&& \int_0^{2 \pi} d(\theta_k - \theta_p )  e^{i 3(\theta_k - \theta_p )} \frac{e^{-i 2 (\theta_k - \theta_p )} - 1}{\lambda - \cos(\theta_k - \theta_p )} = \nonumber \\
&& \int_0^{2 \pi} d(\theta_k - \theta_p )  e^{i 2(\theta_k - \theta_p )} \frac{e^{-i  (\theta_k - \theta_p )} - e^{ i  (\theta_k - \theta_p )}}{\lambda - \cos(\theta_k - \theta_p )} \nonumber \\
&& = 2 I_2 ,
\end{eqnarray}
and similarly for $ l = -1$.
In this way we can get the following expressions for  $ V_{kp}^l , l = 1, 3, -1$,
\begin{equation}
V_{kp}^1 = \frac{2 \pi}{8 \; E_p \; E_k} [ - 2 |m| (\lambda - \sqrt{\lambda^2 - 1})],
\end{equation}

\begin{eqnarray}
 V_{kp}^3 &= & \frac{2 \pi}{8 E_p \; E_k} [ - 2 |m| (\lambda - \sqrt{\lambda^2 - 1})^3  \nonumber \\
&& - \frac{m}{|m|}
\frac{(E_p - m) (E_k - m) (E_p + E_k + 2 m)}{p\; k}   \nonumber \\
&& \times (\lambda - \sqrt{\lambda^2 - 1})^2  ],
\end{eqnarray}
and
\begin{eqnarray}
V_{kp}^{-1} &=& \frac{2 \pi}{8 E_p \; E_k} [ 2 |m| (\lambda - \sqrt{\lambda^2 - 1}) \nonumber \\
&& + \frac{m}{|m|}
\frac{(E_p - m) (E_k - m) (E_p + E_k + 2 m)}{p\; k}  \nonumber \\
&&  \times   (\lambda - \sqrt{\lambda^2 - 1})^2  ].
\end{eqnarray}

Note that we take
\begin{equation}\label{r_def}
 \lambda - \sqrt{\lambda^2 - 1} = \left\{ \begin{array}{cc}
                                            \frac{k}{p}, & k<p \\
                                            \frac{p}{k}, & p<k
                                           \end{array}
\right. \equiv r_{pk}
\end{equation}
as $\sqrt{\lambda^2 - 1} = \sqrt{\frac{k^4+p^4+2k^2p^2}{4k^2p^2} - 1} = \sqrt{\frac{k^4+p^4-2k^2p^2}{4k^2p^2}} = \frac{\sqrt{(k^2-p^2)^2}}{2kp}$. At this point we choose $\sqrt{(k^2-p^2)^2}=|k^2-p^2|$, which then leads to Eq.\ref{r_def}. Other choices lead to an unphysical $V$ that does not decay to zero with large $k$ and $p$ and diverges at $k=0$ or $p=0$.
The general expression for $V$ in the three cases of interest $ l = 1, 3, -1 $ is given by
\begin{eqnarray} \label{final_V_def}
V_{kp}^l &=& \frac{2 \pi}{8 E_p \; E_k} \Bigg[ -2 \,\mathrm{sgn}(l) \,|m|\, r_{kp}^{|l|} \nonumber \\
&& - (1-\delta_{l,1})\mathrm{sgn}(l)\mathrm{sgn}(m) \times \\ \nonumber
&&\;\;\;\;\times\frac{(E_p - m) (E_k - m) (E_p + E_k + 2 m)}{p\; k} \,r_{kp}^2  \Bigg].
\end{eqnarray}
where $\delta_{x,y}$ is the Kronecker delta, equal $1$ when $x=y$ and otherwise 0.
Finally, we need to solve
\begin{equation} \label{final_equation}
|\Delta_{\bf k} | = - \frac{1}{2 \pi} \int_0^\infty \; dp \; p \; V_{kp}^l \frac{|\Delta_{\bf p}|}{{\cal E}_p},
\end{equation}
with $V_{kp}^l$ defined in Eq.\ref{final_V_def}. Note that $|\Delta_{\bf k} |$ only depends on $k$.

\subsection{Ground state energy}
For the BCS ground state $|\Omega\rangle$ for which $ \alpha_{\bf k} |\Omega\rangle = 0 $, after a simple algebra,
we have
\begin{equation}
<\Omega| K_{\rm eff} |\Omega> = - \sum_{\bf k} \frac{{\cal E}_k - \xi_k}{2}.
\end{equation}
To make assessment of the implied ground state energies we first note that
\begin{equation}
 \langle b_{\bf k}^\dagger \rangle  = - \frac{\Delta_{\bf k}^*}{2 \; {\cal E}_k} ,
\end{equation}
and thus
\begin{eqnarray}
&& E_0 = <\Omega| K_{\rm eff} |\Omega> - \sum_{{\bf k},{\bf p}} V_{{\bf k} {\bf p}} <b_{\bf k}^\dagger> <b_{\bf k}> = \nonumber \\
&& - \sum_{\bf k} \frac{{\cal E}_k - (E_k - \mu)}{2} -  \sum_{{\bf k},{\bf p}} V_{{\bf k} {\bf p}} \frac{\Delta_{\bf k}^*}{2 \; {\cal E}_k} \frac{\Delta_{\bf p}}{2 \; {\cal E}_p}.
\end{eqnarray}
In the second term (after the infinite volume limit) we need to integrate over $\theta_k$ and $\theta_p$. Because
$ \Delta_{\bf k}^* \Delta_{\bf p} =  |\Delta_{\bf k}| |\Delta_{\bf p}| e^{i l (\theta_k - \theta_p)} $, a change of
variables, $\theta_{+} = \theta_k + \theta_p $ and $\theta_{-} = \theta_k - \theta_p $, is appropriate to apply. The
function under integral, $ f(\theta_k, \theta_p ) \sim e^{i l (\theta_k - \theta_p)} V_{{\bf k}{\bf p}}$, has a periodicity under translations for (multiples of) $2 \pi $ of
$\theta_k$ and of $\theta_p$. After a short analysis of mappings, we can conclude
\begin{eqnarray}
&& \int_0^{2 \pi} d\theta_k \int_0^{2 \pi} d\theta_p f ( \theta_k , \theta_p ) \nonumber \\
&&= \frac{1}{2} \int_0^{4 \pi} d\theta_+
\int_0^{2 \pi} d\theta_- f(\theta_- )  \nonumber \\
&& = 2 \pi \int_0^{2 \pi} d\theta_- f(\theta_- ).
\end{eqnarray}
Therefore the ground state energy density for a fixed angular momentum $l$ instability,  $ \Delta_{\bf k}^* = |\Delta_{\bf k}| e^{i l \theta_k}$, is
\begin{eqnarray} \label{ground_state_energy_polar}
&& E_0^l = - \frac{1}{(2 \pi)} \int dk \; k \frac{{\cal E}_k - (E_k - \mu)}{2} \nonumber \\
&& -  \frac{1}{(2 \pi)^2}
\int_0^\infty \; dp \; p \; \int_0^\infty \; dk \; k \;
  V_{k p}^l \frac{ |\Delta_{\bf k} |}{2 \; {\cal E}_k} \frac{|\Delta_{\bf p} |}{2 \; {\cal E}_p},\nonumber \\
\end{eqnarray}

\subsection{Proof of symmetry between $l=-1$ and $l=3$ channels} \label{app:V_symmetry}

The two pairing channels corresponding to $l=-1$ and $l=3$ satisfy a symmetry relation
\begin{equation}\label{V_symmetry}
 V^{l=3}_{kp}(m) = V^{l=-1}_{kp}(-m)
\end{equation}
and therefore the solutions for these two channels are equal up to a sign-flip of $m$. Here we present the proof of Eq.\ref{V_symmetry}.

First we note that $E_k(m) = E_k(-m)$, as $E_k=\sqrt{k^2+m^2}$. Therefore, $E_k$ is an implicit function of $|m|$. For the sake of clarity we introduce $A_{kp}(m)\equiv\frac{2\pi}{8E_kE_p}$ and $A_{kp}(m)=A_{kp}(-m)$. We also introduce $a\equiv E_k$, $b\equiv E_p$.
We focus here on the case $k<p$ but an analogous proof can be easily given for the case $k>p$.
\begin{eqnarray}
&& V^{l=3}_{kp, k<p}(m) \\ \nonumber
&& = A_{kp}(m)\Bigg[ -2|m|\frac{k^3}{p^3} \\ \nonumber
&& \;\;\;\;\;- \mathrm{sgn}(m)\frac{(a-m)(b-m)(a+b+2m)}{pk}\frac{k^2}{p^2}\Bigg]
\end{eqnarray}
We now separate the second term into parts which are even and odd with respect to $m$
\begin{eqnarray}
 && (a-m)(b-m)(a+b+2m) \\ \nonumber
 && = (ab - am - bm +m^2)(a+b+2m) \\ \nonumber
 && = a^2b + b^2a + 2mab - ma^2 - mab - 2m^2 a\\ \nonumber
 &&\;\;\;\; - mab -mb^2 -2m^2b + m^2a + m^2b + 2m^3\\ \nonumber
 && = a^2b + b^2a - m^2(a+b) - m(a^2+b^2-2m^2)
\end{eqnarray}
Now we perform a change of variables ${\tilde m} = -m$.
\begin{eqnarray}
 && (a-m)(b-m)(a+b+2m) \\ \nonumber
 && = a^2b + b^2a - {\tilde m}^2(a+b) + {\tilde m}(a^2+b^2-2{\tilde m}^2) \\ \nonumber
 && = (a-{\tilde m})(b-{\tilde m})(a+b+2{\tilde m}) + 2{\tilde m}(a^2+b^2-2{\tilde m}^2)
\end{eqnarray}
We now use $\mathrm{sgn}(x) = -\mathrm{sgn}(-x)$, and $\mathrm{sgn}(x) x = |x|$ to obtain
\begin{eqnarray}
&& V^{l=3}_{kp, k<p}(-{\tilde m}) \\ \nonumber
&& = A_{kp}({\tilde m})\Bigg[ -2|{\tilde m}|\frac{k^3}{p^3} \\ \nonumber
&& \;\;\;\;\;+ \mathrm{sgn}({\tilde m})\frac{(a-{\tilde m})(b-{\tilde m})(a+b+2{\tilde m})}{pk}\frac{k^2}{p^2} \\ \nonumber
&& \;\;\;\;\;+ 2|{\tilde m}|\frac{a^2+b^2-2{\tilde m}^2}{kp}\frac{k^2}{p^2} \Bigg]
\end{eqnarray}
We rewrite the additional term using $k,p$ and ${\tilde m}$
\begin{eqnarray}
 && a^2+b^2-2{\tilde m}^2 \\ \nonumber
 && = k^2+{\tilde m}^2+p^2+{\tilde m}^2 -2{\tilde m}^2 \\ \nonumber
 && = k^2+p^2
\end{eqnarray}
and
\begin{eqnarray}
 \frac{k^2+p^2}{kp}\frac{k^2}{p^2}
  = (k^2+p^2)\frac{k}{p^3}
  = \frac{k^3}{p^3}+\frac{k}{p}
\end{eqnarray}
The $\frac{k^3}{p^3}$ terms cancel and we finally obtain
\begin{eqnarray}
&& V^{l=3}_{kp, k<p}(-{\tilde m}) \\ \nonumber
&& = A_{kp}({\tilde m})\Bigg[ 2|{\tilde m}|\frac{k}{p} \\ \nonumber
&& \;\;\;\;\;\;\ + \mathrm{sgn}({\tilde m})\frac{(a-{\tilde m})(b-{\tilde m})(a+b+2{\tilde m})}{pk}\frac{k^2}{p^2} \Bigg] \\ \nonumber
&& = V^{l=-1}_{kp, k<p}({\tilde m})
\end{eqnarray}
\emph{QED.}

\begin{figure}[ht]
\centering
\includegraphics[width=3.2in]{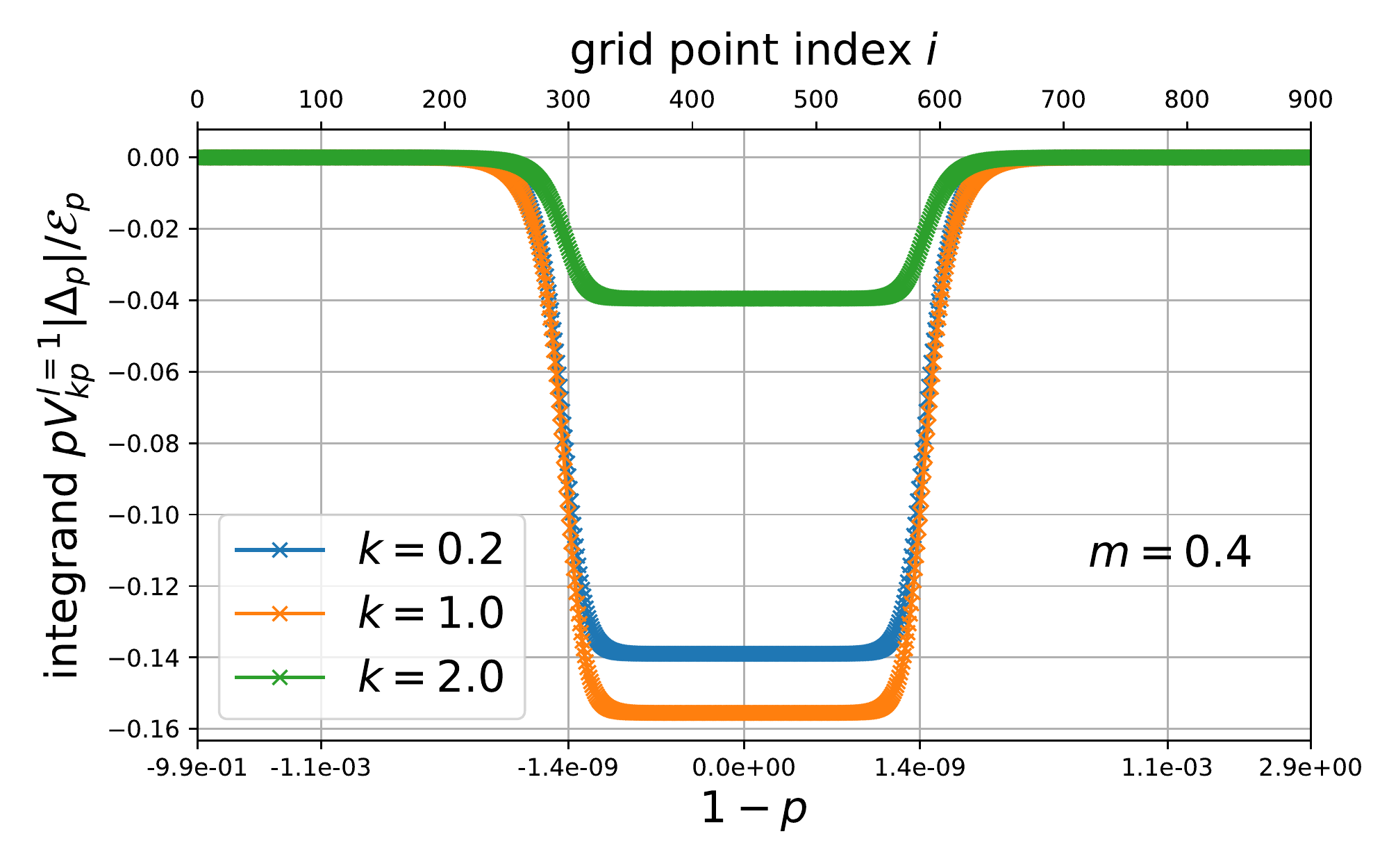}
\caption{Integrand function in the final iteration for $l=1$, $m=0.4$, $k_F=1$. Examples are given for three different $k$. The sharp peak at $k_F$ has a width $\sim 10^{-9}$ and is properly resolved using a logarithmic grid.}
\label{app_fig1}
\end{figure}

\subsection{Numerical solution}
We solve Eq.\ref{final_equation} numerically, using the forward-substitution algorithm. We start from an initial guess for $|\Delta_k|$ (in practice $|\Delta_k|=10^{-5}, \forall k$) and then recalculate it from the RHS of Eq.\ref{final_equation} iteratively until it converges.  We take as the criterion for convergence
\begin{equation}
 \frac{\mathrm{max}_k |\Delta^\mathrm{new}_k - \Delta^\mathrm{old}_k|}{\mathrm{max}_k |\Delta^\mathrm{new}_k|} < 10^{-3} k_F
\end{equation}
It takes 30-130 iterations to satisfy the convergence criterion. We keep $k_F=1$ to set the unit.

We perform the integration on the RHS of Eq.\ref{final_equation} using trapezoid rule. The integrand function on the RHS is very sharply peaked around $k_F$. To properly resolve the integrand function, we disretize $k$ using a logarithmic grid
\begin{equation}
 {\tilde k}_j = e^{a_\mathrm{min}+\frac{j}{N_k}(a_\mathrm{max}-a_\mathrm{min})}, \;\; j\in[0,N_k)
\end{equation}
with $N_k=500$, $a_\mathrm{min}=-30$, $a_\mathrm{max}=4$. The logarithmic grid is placed on both sides of $k_F$, to include all points given by
$$1\pm {\tilde k}_j>0.$$
We add the $k=1$ point by hand. Therefore, our grid can resolve peaks at $k_F$ which have a width $\gtrsim e^{-30}$, which is near the limitation of double precision numeric type. Logarithmic grid is particularly important for the $l=1$ case at low $m$, where the integrand is most sharply peaked. We illustrate our grid and the integrand function on Fig.\ref{app_fig1} We have checked that the results do not depend on the numerical parameters.

\end{document}